\documentclass[aps,pra,reprint,superscriptaddress,10pt,longbibliography]{revtex4-1}
\usepackage[utf8]{inputenc}

\usepackage[T1]{fontenc}
\usepackage{newtxmath,newtxtext}

\usepackage{graphicx}
\usepackage{amsmath}
\usepackage{amsfonts}
\usepackage{bm}
\usepackage{bbm}
\usepackage[mathcal]{euscript}

\usepackage{multirow}

\usepackage{color}
\usepackage{soul}

\usepackage{tikz}
\usetikzlibrary{patterns}

\newcommand{\etal}{\textit{et al}.}

\begin{document}
\title{Machine-learning-driven simulated deposition of carbon films:\\
from low-density to diamond-like amorphous carbon}

\author{Miguel~A. Caro}
\email{mcaroba@gmail.com}
\affiliation{Department of Electrical Engineering and Automation,
Aalto University, Espoo, Finland}
\affiliation{Department of Applied Physics, Aalto University, Espoo, Finland}
\author{G\'abor Cs\'anyi}
\affiliation{Engineering Laboratory, University of Cambridge,
Cambridge CB2 1PZ, United Kingdom}
\author{Tomi Laurila}
\affiliation{Department of Electrical Engineering and Automation,
Aalto University, Espoo, Finland}

\author{Volker L. Deringer}
\affiliation{Department of Chemistry, University of Oxford,
Oxford OX1 3QR, United Kingdom}

\newcommand{\eq}[1]{Eq.~(\ref{#1})}
\newcommand{\fig}[1]{Fig.~\ref{#1}}

\begin{abstract}
Amorphous carbon (a-C) materials have diverse interesting and useful properties,
but the understanding of their atomic-scale structures is still incomplete.
Here, we report on extensive atomistic simulations of the deposition and growth of a-C films,
describing interatomic interactions using a machine learning (ML) based Gaussian Approximation
Potential (GAP) model. We expand widely on our initial work [Phys. Rev. Lett. {\bf 120}, 166101
(2018)] by now considering a broad range of incident ion energies, thus modeling samples that
span the entire range from low-density ($sp^{2}$-rich) to high-density ($sp^{3}$-rich,
``diamond-like'') amorphous forms of carbon. Two different mechanisms are observed in these
simulations, depending on the impact energy: low-energy impacts induce $sp$- and $sp^{2}$-dominated
growth directly around the impact site, whereas high-energy impacts induce peening. 
Furthermore, we propose and apply a scheme for computing the anisotropic elastic properties of the
a-C films. Our work provides fundamental insight into this intriguing class of disordered solids,
as well as a conceptual and methodological blueprint for simulating the atomic-scale deposition of
other materials with ML-driven molecular dynamics.
\end{abstract}

\date{\today}

\maketitle

\section{Introduction}

Since the early days of molecular dynamics (MD) simulations and materials modeling,
carbon has received intense attention, given its importance in organic compounds but
also in elemental forms. Besides the crystalline phases of pure carbon,
some of which possess mechanical and electronic properties unmatched by
any other compound, the complex and density-dependent structures and properties of
amorphous carbon (a-C) have also been reported and exploited~\cite{robertson_2002}.
The ability of carbon to form diverse structural environments and chemical bonds
has long been a challenge for simulations, requiring highly flexible \textit{and}
accurate interatomic potentials. Many efforts have been devoted to the development
of potentials for the study of nanoscale allotropes of carbon, including ``diamond-like''
or tetrahedral amorphous carbon (ta-C) 
\cite{tersoff_1988, brenner_1990, marks_2000, los_2003, erhart_2005, pastewka_2008, srinivasan_2015}.

Molecular dynamics studies of materials have traditionally been done with empirically
fitted interatomic potentials of relatively simple functional form~\cite{keating_1966},
typically containing harmonic terms and two- and three-body interactions (distances
and angles) only. MD simulation has now become a
popular tool routinely used in physics, chemistry, materials science, and molecular biology
to study complex systems at the atomic scale. Still, accuracy remains an issue, even for
the best empirical potentials currently available.
To reliably handle bond breaking and highly anharmonic potential
energy surfaces, one must often rely on {\em ``ab initio''} MD methods, typically based on
density-functional theory (DFT). Unfortunately, DFT-MD simulations are several orders
of magnitude more costly than classical MD, severely limiting the accessible
system sizes and time scales. Carbon is a prime example: different routes to
computationally generate ta-C structures have been explored in detail, the
most popular being the ``liquid quench'' technique
\cite{galli_1989, marks_1996, mcculloch_2000, marks_2002, risplendi_2014, ranganathan_2017, jana_2019}.
Explicit deposition of carbon atoms~\cite{kaukonen_1992, uhlmann_1998, jaeger_2000, gao_2003, marks_2005,
moseler_2005, li_2013c, caro_2018, liu_2019, wang_2020}, 
mimicking ta-C film growth under experimental conditions,
is too computationally costly to
be practical at the DFT level. Alternative generation techniques, including quenching from the
simulated melt, invariably fall short, each to a different extent, of predicting experimental
$sp^3$ values~\cite{laurila_2017}, which can be as high as 90\% for ``superhard''
ta-C~\cite{schultrich_1998}.

We have recently shown that this problem can be overcome by using a machine learning (ML)
based interatomic potential~\cite{caro_2018}, which provides close to DFT-level accuracy
and flexibility at a small fraction of the cost. We showed that
explicit deposition of ta-C, simulated within the Gaussian approximation potential (GAP)
framework~\cite{bartok_2010} using the 2017 GAP for carbon~\cite{deringer_2017}, provides a
satisfactory description of the structural properties observed experimentally and also
insight into the microscopic growth mechanism of ta-C~\cite{caro_2018}.
We review the salient aspects of ML-driven simulations below, and we mention in passing
earlier studies of crystalline carbon with such potentials, which described the graphite--diamond
coexistence \cite{khaliullin_2010} and a transformation mechanism between the two allotropes
\cite{khaliullin_2011}.

In the present work, we use large-scale ML-driven atomistic simulations to generate a-C 
films over the full range of mass densities. We thereby extend and complement our earlier
work which focused on high-density ta-C films \cite{caro_2018}, and we obtain more general
and systematic insight into the structures and properties of amorphous forms of carbon,
including low density films and their surface properties. 
This study covers relevant structural and mechanical properties,
an elucidation of the growth mechanism, and the dependence of all these properties on
deposition energy and mass density. In addition to this fundamental insight, we provide a
comprehensive dataset of atomistic structures to enable future work in the field.

\section{Methodology}

\subsection{Gaussian approximation potential (GAP) modeling of amorphous carbon}

The Gaussian approximation potential (GAP) framework is an ML approach to generating
interatomic potentials, performing a high-dimensional fit to reference quantum-mechanical
data~\cite{bartok_2010}. 
Such ML-based potentials bring large system sizes and long MD trajectories within reach, while
(largely) retaining the accuracy of the underlying reference data.
Overviews of these emerging methods are found, e.g., in 
Refs.~\cite{behler_2017, deringer_2019, Noe2020}. 
In the GAP framework, similarity functions or \textit{kernels} are used to quantify
how similar an atom in a candidate structure is to another atom in the reference
database~\cite{bartok_2013, bartok_2015}. Here, we use a GAP model 
that was developed specifically with liquid and
amorphous carbon in mind: most structures in the reference database, therefore,
are snapshots from DFT-MD or GAP-MD simulations of those disordered phases. 
The resulting potential has been
validated, initially, for structural and mechanical properties of the bulk, for
surface energies and reconstructions~\cite{deringer_2017,deringer_2018}, and for porous
($sp^{2}$-rich) carbon materials at lower densities as used in energy storage \cite{deringer_2018a}.

A special requirement for deposition simulations, in which high-energy impact events lead to
locally strongly disordered structures, is that the potential must be highly flexible. 
This is critical as structures from actual deposition simulations cannot serve for the iterative
generation of reference data directly (they are out of reach even for single-point DFT evaluations). 
Recent evidence suggests that GAPs can be made flexible
enough to provide a physically meaningful representation of potential-energy surfaces both in
the low- and higher-energy regions.
For example, they have been coupled to crystal-structure searching, in which structures
``unknown'' to the potential can be identified in an ML-driven search, 
initially demonstrated for the carbon GAP \cite{deringer_2017b}. 
Together with the previously evidenced high quality of the deposition simulations, i.e., the
good agreement with experimental observables observed in initial work \cite{caro_2018}, this
suggests that the carbon GAP is indeed able to capture the deposition process correctly.
In this context, we mention the recently demonstrated usefulness of GAP simulations for radiation
damage in elemental tungsten and silicon, where the impact of (very) highly energetic ions must be
correctly described as well \cite{byggmaestar2019, DominguezGutierrez2020, Hamedani2020}.

\subsection{GAP-driven deposition simulations}

\subsubsection{Simulation protocol}

\begin{figure}[t]
\includegraphics{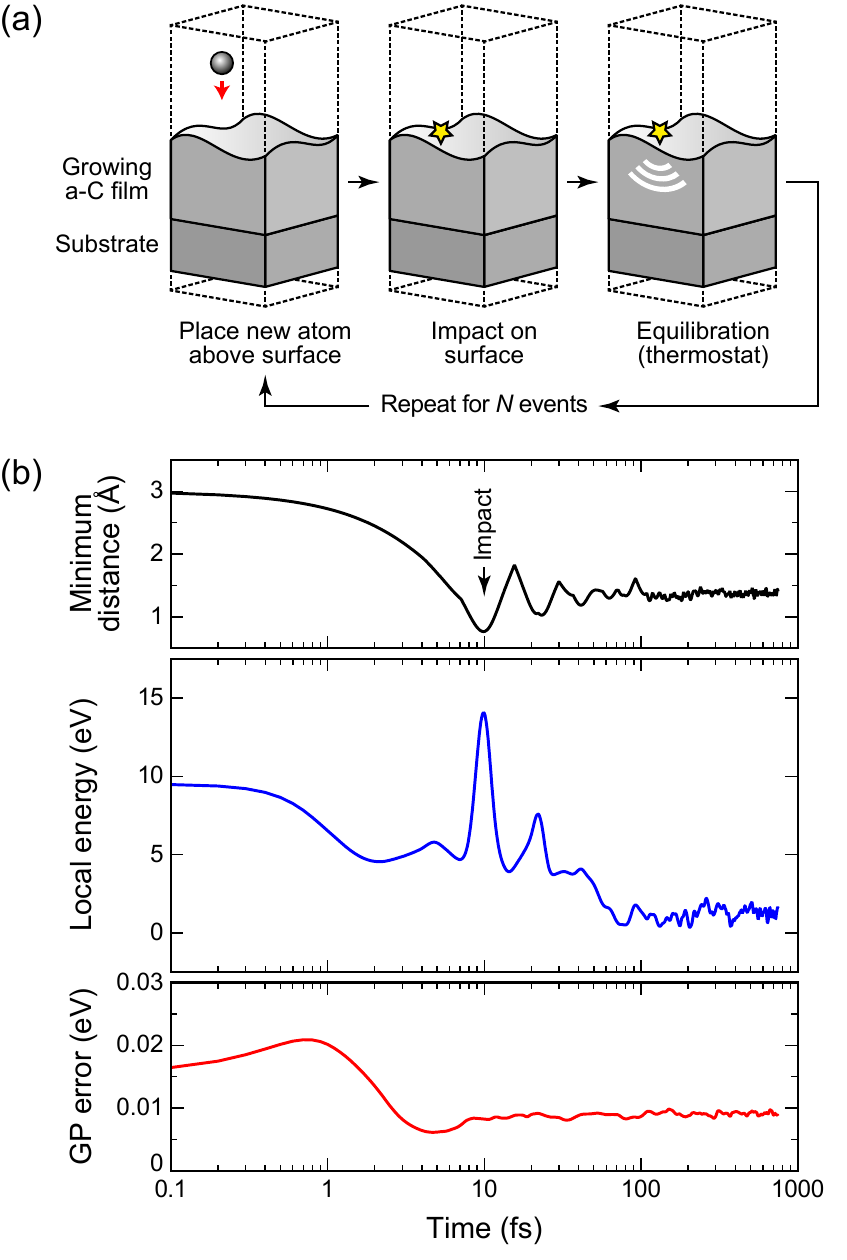}
\caption{Modeling amorphous carbon (a-C) film growth by deposition simulations.
(a) Schematic of the computational protocol. A carbon atom is randomly placed above the surface
and obtains an initial velocity corresponding to a given energy (between 1 and 100 eV). The atom
impacts the surface after about 10 fs of simulation time, and the system is then thermostatted
for several hundred fs (up to 1~ps), before the next deposition event takes place. Dashed lines
indicate the periodic boundaries of the simulation cell. (b) A selected single 60 eV deposition event,
characterized using properties of the impacting atom as described by the GAP. 
{\em Top:} Distance of the impacting atom from its respective closest neighbor. The atom is
initially placed at 3 \AA{} above the surface and quickly approaches it (note the logarithmic
scale of the
horizontal axis); the shortest C--C contact formed by this atom (below 1~\AA{}) is registered
10~fs after the event has started, and it then settles in at an interatomic distance of
$\approx 1.4$~\AA{}, in line with the values for diamond and graphite. {\em Middle:} GAP local
energy of the impacting atom, showing a spike upon impact (consistent with the smallest nearest-neighbor
spacing at around 10 fs of simulation time), and then a settling-in of the energy slightly above that
of ideal diamond (which is set as the energy zero) as the local
environment of the atom relaxes. {\em Bottom:} Predicted error of the Gaussian process (GP), used here to
quantify the error of the prediction in the sense of how far the local environment of the
incident atom is away from those described by the reference database. 
}
\label{plot_full_event}
\end{figure}

The methodology used to generate high-density ta-C films was outlined in our initial
work~\cite{caro_2018}, and it is sketched in Fig.\ \ref{plot_full_event} (a) in a simplified
way. In the present section, we expand significantly on prior work by discussing
error estimates for the GAP prediction and the
nature of overcoordinated carbon atoms. 
Moreover, the protocol to carry out the deposition simulations
is described here in full detail for consistency.

\cleardoublepage 
%% This cleardoublepage is needed to place the figure where we want it %%

Simulated deposition of a-C was carried out starting with a (111)-oriented diamond slab with
3240 C atoms in periodic boundary conditions
(PBC) as substrate. The stable $2 \times 1$ surface reconstruction
was used to avoid the presence of highly energetic dangling bonds at the top and bottom 
of the slab. The role of initial substrate size on growth is briefly discussed in the
Supplemental Material (SM)~\footnote{See Supplemental Material at [URL will be
inserted by publisher] for miscellaneous technical detail about the simulations:
role of substrate in initiating growth, time steps used and thermostat effects, including
visualization of the thermal spike upon impact. We also provide mass density profiles for
all the films and more detail on the COHP and COOP bonding analysis of 5-c complexes.}.
We then generated an a-C template by depositing 2500 C atoms
with kinetic energy of 60~eV
onto the diamond substrate. Afterwards, this template was used as substrate for all subsequent
deposition simulations in the energy regimes that we explored, viz. between 1 and 100~eV. An
additional 5500 C atoms were deposited at the chosen deposition energy. The initial position of each
incident atom was randomly chosen within the $xy$ plane of the simulation box; the initial $z$
coordinate was chosen so that the incident atom was at least 3~{\AA} away from the first atom that
it found on top of the film in its downward trajectory within a cylinder of radius 1~{\AA}
(Fig.~\ref{plot_full_event}). After impact, most incident atoms were
observed to predominantly deposit into the film by bonding to the substrate. Rarely, the incident
atom bounced off (determined according to a connectivity criterion), 
in which cases the deposition event was repeated with different initial
conditions. Occasionally, small portions of the growing films detached after the impact,
resulting in groups of atoms ``floating'' in the simulation box. Those atoms were removed
from the simulation box before the system was prepared for the following deposition event.

\begin{table}[t]
\caption{Protocol for simulating a single impact event (at 60 eV): the time step is small at
first, and then is gradually increased once the impacting atom ``settles'' in the slab. Settings
for other energies are given as Supplemental Material.}
\begin{ruledtabular}
\begin{tabular}{r r r}
Time step & Number of steps & Time \\
\hline
0.1~fs & 200 & 20~fs \\
0.25~fs & 120 & 30~fs \\
0.5~fs & 100 & 50~fs \\
1~fs & 200 & 200~fs \\
2~fs & 225 & 450~fs \\
\hline
 Total & 845 & 750~fs \\
\end{tabular}
\end{ruledtabular}
\label{01}
\end{table}

In all cases, the substrate temperature was kept fixed at $\approx$ 300~K using LAMMPS's
implementation of the Nos\'e-Hoover thermostat~\cite{nose_1984,hoover_1985,parrinello_1981,
martyna_1994,shinoda_2004,tuckerman_2006}. Each
impact event itself, which consisted of the first few fs of dynamics, was run in the $NVE$ ensemble;
after that, the thermostat, with time constant of 0.1~ps, was switched on and the MD was run in the
$NVT$ ensemble until equilibration was reached. The required equilibration time depended strongly on
the kinetic energy of the incident ion, since this value determined the amount of excess kinetic
energy which needs to be removed. To avoid excessive CPU costs, we optimized MD time steps and
equilibration times for each deposition regime, following the general guideline that atomic positions
should not change by more than 0.1~{\AA} per time step. A representative example is given in
Table~\ref{01}; more detailed information regarding the choice of time steps is provided in the SM.

The choice of thermostat for this kind of simulation is not straightforward. In this work we
settle for applying the thermostat to all atoms, as opposed to applying a ``wall'' thermostat,
as done, e.g., by Marks in previous ta-C deposition simulations~\cite{marks_2005}.
Under periodic boundary conditions, how the excess kinetic energy is removed from
the supercell (to bring it back to its nominal temperature) is problematic. There is no
simple solution to that problem since either 1) some kinetic energy will be recycled
through the periodic boundaries (the present case) or 2) unrealistic dynamics will be
enforced by introducing a wall thermostat that acts as a heat sink, where the inner atoms
are not coupled to the thermostat. The best solution to the problem is indeed making the system
so large that the role of the thermostat becomes secondary, at the expense of
the associated increase in computational
cost. Another consideration, applicable to a-C in particular, is that thermal transport
is hindered compared to, say, graphene, due to the
disordered atomic structure. In the SM we show that, even at 100~eV, the highest deposition
energy studied here, the thermal spike upon impact is fairly localized in comparison to the
dimensions of the supercell, removing the need for wall thermostats. We also show that the choice
of coupling constant is sensible, within the context of how long it takes to relax
the global temperature increase induced by the thermal spike. Videos characterizing
the thermal spike following a 100~eV deposition event can be retrieved from Ref.~\cite{caro_2020b}.

To model the atomic interactions, we used a GAP optimized for a-C~\cite{deringer_2017}. Detailed
analyses of structural and elastic features of the deposited films were performed for all
structures. We used LAMMPS for all deposition simulations~\cite{plimpton_1995,ref_lammps}. For
visualization, structure manipulation, etc., we used ASE~\cite{larsen_2017},
VMD~\cite{humphrey_1996,ref_vmd,kohlmeyer_2017}, OVITO~\cite{stukowski_2010}, and different
in-house codes, some of which are publicly available~\cite{ref_deposition}.

\begin{figure}[t]
\includegraphics{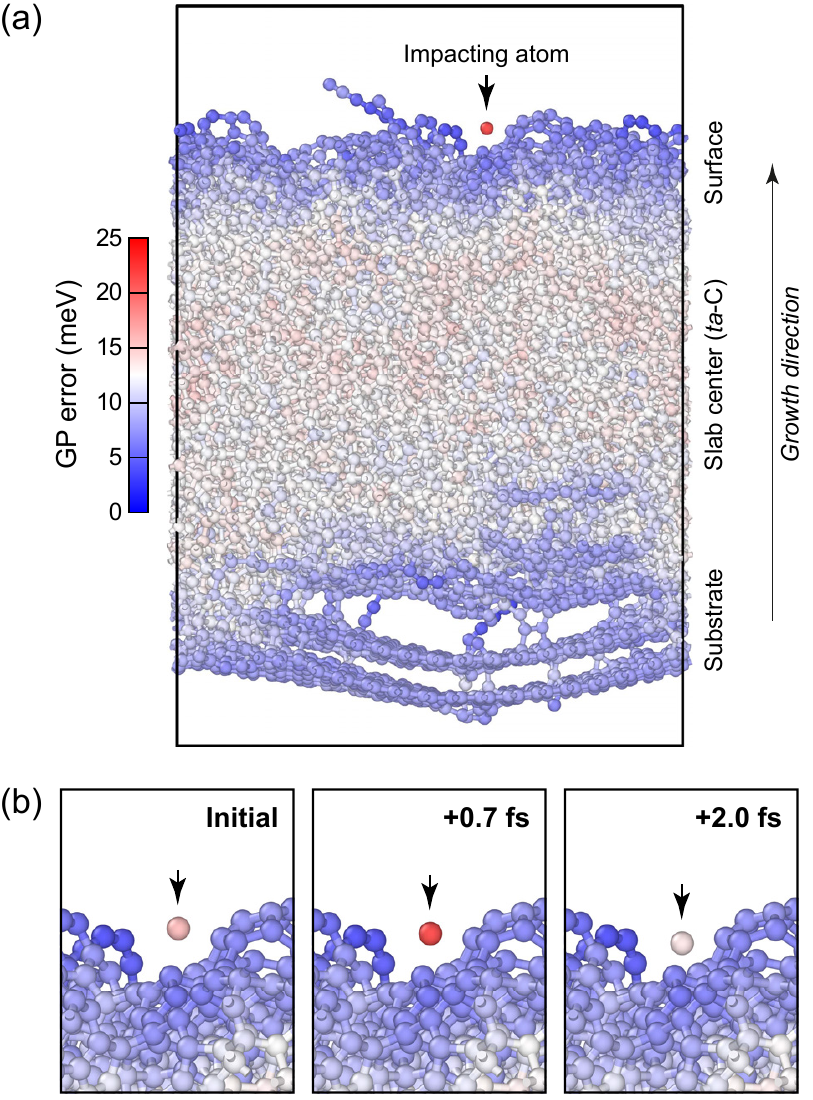}
\caption{Uncertainty quantification through the predicted GP error \cite{bartok_2018}, 
as in \fig{plot_full_event} (c),
but now indicating the per-atom error for each individual atom in a slab by color-coding 
(in a style similar to Ref.~\onlinecite{bartok_2018}). As a qualitative rule of thumb, atoms in 
blue correspond to configurations which are similar to those contained in the GAP fitting database
\cite{deringer_2017}, whilst those in red correspond to configurations which are further away. 
(a) Side view of the ta-C slab during the
test deposition event characterized in \fig{plot_full_event} (b), at +0.7 fs into the
simulation (i.e., at the point where the GP error reaches its maximum for the impacting atom).
(b) Close-up of the impact region at the beginning of the specific event (``initial'') and at
+0.7 and +2.0 fs into the simulation, respectively. This sequence shows, again complementing
Fig.\ \ref{plot_full_event} (b), that after only 2.0 fs the incident atom no longer
has an unusually large GP error compared to other atoms in the center of the slab.
}
\label{fig_local_gap_error}
\end{figure}

\subsubsection{Error estimates}

During our deposition simulations, impinging atoms experience highly energetic,
off-equilibrium configurations. Since the systems contain thousands of atoms, it is not feasible
to compute DFT reference data for such systems and to feed them into the GAP fitting database; instead, the
potential has to make predictions based on existing data for small systems. It is therefore important to
determine how representative the reference data are. For this, we use
the intrinsic uncertainty estimation of the underlying Gaussian process to determine the expected
error of a prediction for any given atomic environment. The variance of the GP prediction (which
has a dimension of energy squared) is taken to be the square of the prediction error, as discussed in
Ref.~\onlinecite{bartok_2018}. We determine this quantity along a separate test trajectory in which we
sampled {\em all}
individual MD steps for a few impact events. At each step, the variance of the prediction is
obtained for each individual atom, and we focus on the incident one for now. 
Our analysis [Fig.\ \ref{plot_full_event}~(b)] shows that even during
the impact itself, when the atom comes closer than 1 \AA{} to its nearest neighbor (corresponding to a
bond compression of almost one-third compared to equilibrium), the error of the
prediction is in the region of 10--20 meV/atom. 

For a more comprehensive view, we color-code all atoms in a given slab by the GP predicted
error, as shown in \fig{fig_local_gap_error}.
The overview figure in panel (a) provides general insight into the slab: 
the bottom region is presumably well represented in the reference database, but so is the $sp^{2}$-rich 
surface region. This reflects the fact that the potential is {\em explicitly fitted} to small surface
slabs including strongly disordered configurations. The region where the predicted error is higher, 
although not extremely high, is the center of the slab. Again, this can be understood because most 
reference data are from iterative GAP-MD quenches, and typically reach 60-70\% $sp^{3}$ count
\cite{deringer_2017}, but not the 90\% that are characteristic of the dense regions in our
as-deposited slabs \cite{caro_2018}.
The fact that, despite the residual GP error in this region, we are nonetheless observing a structure 
which is consistent with experiment \cite{caro_2018} suggests that the GP variance in the present
simulations is at an acceptable level, and that it does not yet correspond to a region of
configuration space where there is strong extrapolation, at least for this specific system.

Figure \ref{fig_local_gap_error}~(b) offers three close-ups: at the start, at the point of
highest GP variance for the impacting atom [the maximum in \fig{fig_local_gap_error}~(b)],
and then just 1.3~fs
later when the atom is approaching the surface and becoming more similar to structures
which the potential has previously ``seen'' (i.e., which were included as part of the
training set~\cite{deringer_2017}). Summarizing, the GP error analysis provides support for a
reliable description by our GAP model of the physical processes involved in a-C growth,
consistent with the observation of an $sp^{3}$ count in agreement with experiment~\cite{caro_2018}.

\begin{figure*}[t]
\includegraphics{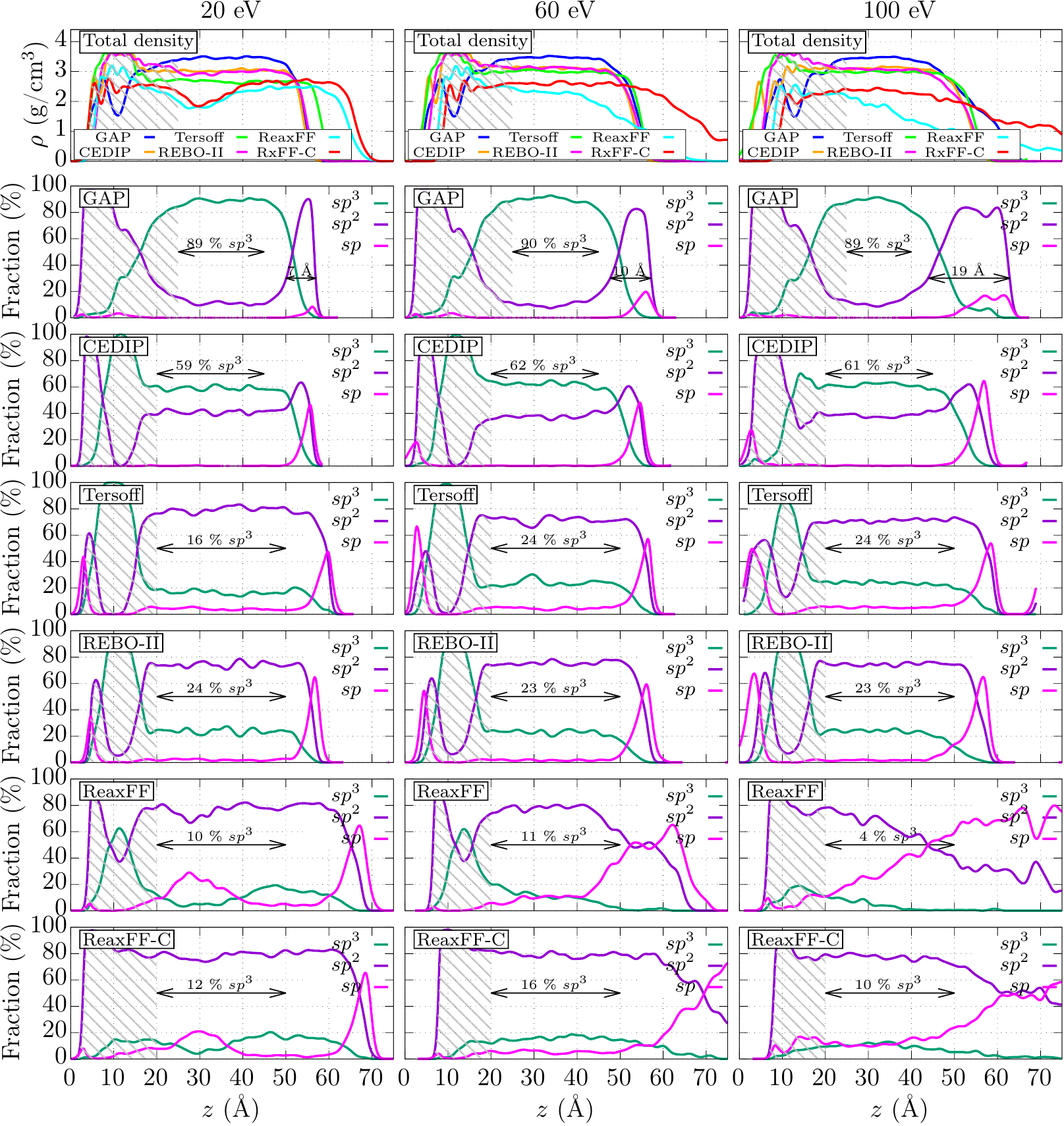}
\caption{Mass density profiles and coordination fractions
(based on a cutoff radius of 1.9~{\AA} for nearest neighbors) for the high-density ta-C films in the
range 20--100~eV, generated using the simulated deposition protocol described in the main
text. We repeated the GAP depositions reported in our earlier work~\cite{caro_2018}, from which
data are plotted here, with five
other popular interatomic potentials for carbon, namely CEDIP~\cite{marks_2000},
Tersoff's potential as parametrized by Erhart and Albe~\cite{erhart_2005},
REBO-II~\cite{brenner_2002}, and two versions of ReaxFF~\cite{vanduin_2001}:
one optimized for hydrocarbons and carbohydrates (ReaxFF)~\cite{chenoweth_2008}
and one optimized for pure carbon (ReaxFF-C)~\cite{srinivasan_2015}.
The GAP succeeds at reproducing experimental high densities and $sp^3$ fractions, and
also the same evolution of surface morphology with deposition energy
as observed experimentally~\cite{davis_1998}: in the ta-C regime, the bulk properties of the
film remain constant (circa 90~\% $sp^3$ bonding) but the width of the $sp^2$-rich surface
region increases monotonically with energy. This subtle feature is not
observed with the other potentials. The shaded areas indicate the
portion of the films corresponding to the initial substrate (the same substrate is used for all
deposition energies, see text).}
\label{06}
\end{figure*}

\subsubsection{Comparison with empirical interatomic potentials}\label{27}

In \fig{06} we show a comparison of our ta-C (i.e., high-density) films reported in
Ref.~\cite{caro_2018} with the outcome of five selected,
empirical reactive potentials for carbon (including a-C). Specifically, we carried out
deposition simulations using:
1) the environment-dependent interaction potential for carbon (CEDIP)~\cite{marks_2000};
2) the Tersoff potential~\cite{tersoff_1988} as parametrized by Erhart and Albe~\cite{erhart_2005};
3) the second-generation reactive empirical bond-order potential (REBO-II) of Brenner
\etal~\cite{brenner_2002};
4) the reactive force field (ReaxFF) of van Duin~\cite{vanduin_2001} as optimized
and parametrized for hydrocarbons and carbohydrates~\cite{chenoweth_2008}; and
5) a ReaxFF parametrized and optimized for pure carbon (ReaxFF-C)~\cite{srinivasan_2015}.
These classical potentials are commonly used for MD simulations of
large systems, and have been recently reviewed by de~Tomas
\textit{et al}.~\cite{deTomas_2016}. 
In this context, we may refer the reader to critical discussions of technical aspects
of empirical potentials \cite{Pastewka2012, Tangarife2019}, and to a benchmark study of various
such potentials specifically with a view to quantify their performance for amorphous
forms of carbon \cite{detomas_2019}. We emphasize that a similar benchmark of
many empirical potentials is outside the scope of the present work.

To characterize atomic coordination environments, as is commonly done in the literature, we count
the number of
neighbor atoms within a sphere, whose radius is chosen at the first minimum of the radial distribution
function~\cite{marks_1996,caro_2014,marks_2000} of a-C, corresponding to the boundary between
first- and second-nearest neighbor shells. We choose the cutoff distance as
1.9~{\AA}~\cite{caro_2014}, with values of 1.85~{\AA} also common in the literature (this distance
corresponds to the minimum of the radial distribution function, and therefore the coordination
counts change very little between 1.85 and 1.9~{\AA}). The assignments are of $sp$, $sp^2$
and $sp^3$ hybridizations for C atoms with 2, 3 and 4 neighbors, respectively.

The failure of a particular potential
to simulate ta-C growth by deposition (defined as leading to an $sp^{3}$ count that strongly deviates from 
experiment, seen most notably for ReaxFF in \fig{06}) certainly points to an
existing deficiency in the force field but does not necessarily mean that it will not perform
well for a different problem (e.g., the graphitization simulations studied in
Ref.~\onlinecite{detomas_2019}). In particular, the limitation of ReaxFF with respect to deposition
simulations can be traced back to the lack of explicit inclusion of exchange repulsion. In
Appendix~\ref{23} we show a more general comparison of these force fields to predict energies
for a database of a-C structures~\cite{deringer_2017}. Even though ReaxFF-C shows very accurate
predictions for most of the structures in the database, it fails to accurately predict the correct
form of the PES for the dimer dissociation curve at close interatomic separations. Therefore,
we refrain from making a \textit{general} assessment of the quality of the different force fields
compared here, especially given that deposition (and other high energy events, e.g., pertaining to
radiation damage) is a very specific type of atomistic simulation.

While the GAP manages to correctly
reproduce the high $sp^3$ fractions observed experimentally~\cite{caro_2018},
together with the deposition energy dependence of the width of the $sp^2$-rich surface
region~\cite{davis_1998}, the other potentials
are unable to achieve these numbers. In particular, the Tersoff potential and the
similarly performing REBO-II severely underestimate the
amount of $sp^3$-bonded carbon for the range of energies under study, while both
versions of ReaxFF predict extremely low $sp^3$ concentrations. In fact, for the higher deposition
energies (60~eV and 100~eV), it was difficult to get the ReaxFF ta-C films to grow at all: portions
of the surface routinely detached from the rest of the film.
Another feature of Tersoff, REBO-II and CEDIP simulations is the existence of 
significant amounts of $sp$ carbon right at the surface, whereas for GAP the amount of
observed $sp$ carbon is much lower. ReaxFF exacerbates this artifact for the
high-energy deposition simulations, where $sp$-bonded carbon is the predominant surface
atomic motif. We reiterate that the issue with ReaxFF can be traced back to the lack
of explicit repulsion interaction, which specifically affects deposition simulations (Appendix~\ref{23}).

A critical practical
point to raise here is that the improved accuracy of GAP does not come ``for free''. Indeed,
a GAP MD simulation is significantly more CPU expensive to run than CEDIP/ReaxFF (which are both 1-2 orders of
magnitude cheaper than GAP) or Tersoff/REBO-II (2-3 orders of magnitude cheaper)~\cite{detomas_2019}. However,
for accuracy-critical applications where the only previous option was to run DFT simulations,
GAPs and other ML-based interatomic potentials offer the capability to run simulations at
similar accuracy but being orders of magnitude cheaper than DFT. In addition, current ongoing
efforts are expected to deliver an order of magnitude speedup for GAP potentials in the near
future~\cite{caro_2019}.

To find the root of the discrepancy between Tersoff, REBO-II, ReaxFF, CEDIP and GAP results, we
give a brief description of these potentials.
The Tersoff potential, the first bond-order potential to be introduced,
consists of a combination of attractive and repulsive pair-wise
interactions, as in Lennard-Jones or Morse potentials, which are switched on or off based
on a smooth cutoff function (the interactions are usually restricted to the first-neighbors
shell). In Tersoff's approach, the attractive potential is scaled by a bond-order (environment-dependent)
parameter which, for carbon, favors 3- and 4-fold coordinations in honeycomb and tetrahedral
configurations, respectively. The REBO-II potential is almost identical to Tersoff, with
modified analytical expressions for the pair-wise interactions. CEDIP works similarly, but
incorporates the knowledge about the atomic coordination explicitly into the form of
the potential. This makes CEDIP more accurate and flexible than Tersoff and REBO-II~\cite{deTomas_2016},
but also significantly more expensive to run. Finally, ReaxFF (see Ref.~\cite{senftle_2016}
for a recent review) introduces high flexibility and numerous terms, including terms for explicit
treatment of dispersion and electrostatics. This means that each potential has its own
range of applicability, with CEDIP and ReaxFF
being used for medium-to-large systems where accurate characterization of structural
transitions and the effect of temperature are required~\cite{suarez_2012}, whereas Tersoff
and REBO-II (and similar potentials, such as that by Brenner~\cite{brenner_1990}) are used to study
very large systems with up to millions of atoms and long time scales~\cite{krasheninnikov_2010}.

A fundamental difference between these potentials and GAP is the introduction in the former
of analytical constraints motivated by observed chemical trends. Namely, the analytical
form of Tersoff, REBO-II and CEDIP potentials gives preference to 3- and 4-fold coordinated complexes
in carbon materials because stable forms of carbon (e.g., graphite and diamond, respectively)
are observed to display such trends. These constraints enable a more accurate description of
the potential energy surface around equilibrium, but possibly at the cost of penalizing
high-energy complexes with non-standard coordinations, as we will see in the next section.
This can manifest itself in the form
of severely overestimated free energy barriers along the paths connecting metastable states.
Such an analysis could explain why CEDIP and, especially, the Tersoff and REBO-II potentials
fail at transforming $sp^2$
carbon into $sp^3$ carbon, a process which is critical for the formation of ta-C and that
will be discussed in detail in the remainder of this manuscript. In stark contrast,
the GAP is designed to reproduce \textit{the data}, with
no physico-chemical constraints other than the assumption of locality and smoothness
for the potential
energy surface. Therefore, there is no fundamental reason why a GAP could not predict the
potential energy surface in the vicinity of highly unstable complexes as accurately as
equilibrium structures, as long as the required data are available. 
This in turn leads to improved estimates of free energy barriers
connecting metastable states, in particular for the case at hand, viz. highly disordered $sp^2$
and $sp^3$ carbon environments.

\subsubsection{Overcoordinated atoms}

We observed that a small but non-negligible number of atoms acquired 5-fold coordination during
the deposition of the high-density samples (that is, five neighbors at distances shorter than
1.9 \AA{} around a single atom). This issue has also been recently 
highlighted in Ref.~\cite{detomas_2019}. Five-fold coordinated C atoms are considered to be
coordination defects,
therefore they are highly energetic and one should expect them to not be present in significant
numbers in the generated structures. The presence of such atoms must be further analyzed since
it could be indicative of an artifact of the potential.
When looking in detail at the statistics,
we find that indeed the number of 5-fold coordinated C atoms is very small. For example,
1.2\% of \textit{deposited} C atoms (that is, excluding the substrate atoms) in the last
snapshot of the 60~eV deposition are 5-fold coordinated. Compare this to the 1.7\%
figure for 5-fold coordinated \textit{incident} C atoms. This means that 29\% of atoms
which were deposited with
5-fold coordination moved away from that configuration into a more stable one as the
simulation progressed.

To understand why the other 71\% remain 5-fold-coordinated,
one needs to note that coordination is conventionally computed based on a nearest-neighbor cutoff
distance~\cite{marks_1996,caro_2014,marks_2000}; even in the context of DFT-based studies,
$sp^2$ vs. $sp^3$ character is sometimes based on a cutoff criterion.
We choose the cutoff distance as 1.9~{\AA}, the location of the minimum between
the first and second peaks in the radial distribution function. A way to determine that
5-fold-coordinated (``5-c'') atoms are not an artifact of the potential is to look at the distance
distribution of neighbors for those atoms: if there were 4 neighbors at distances close
to that of diamond (around 1.5~{\AA}) and another considerably further away (say, 1.8~{\AA}),
that would suggest that the first 4 atoms contribute much more strongly to the bonding than the fifth. 
For the 60~eV deposition \cite{caro_2018}, 
the results for average distances from closest to furthest neighbors are
$(1.462, 1.505, 1.546, 1.602, 1.756)$, in {\AA}. 
As expected, the 5th neighbor is on average significantly further away than the other ones:
only 0.05\% of all
atoms in our 60~eV film had 5 neighbors all closer than 1.6~{\AA}. 

\begin{figure}[t]
\includegraphics{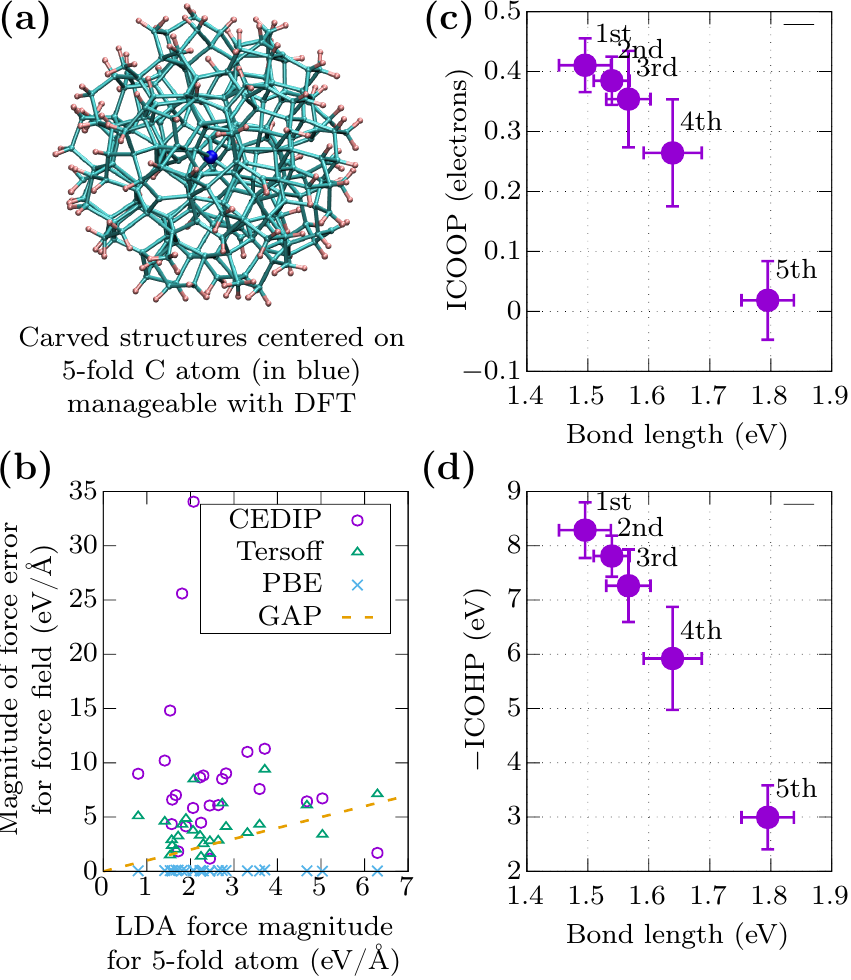}
\caption{Analysis of the 5-fold coordinated (5-c) atomic environments that were occasionally
observed in GAP deposition simulations. 
(a) To make the DFT calculations computationally manageable, 25 spherical
structures centered on each of the 5-c atoms in question, 7~{\AA} in radius (containing
circa 200-250 C atoms), were carved out of the melt-quench GAP structure obtained from the
authors of Ref.~\onlinecite{detomas_2019}; the passivation of the outer C atoms with H followed
the recipe presented in Ref.~\onlinecite{caro_2014}. (b) Comparison of the force
acting on the central 5-c atom predicted at the LDA-DFT level of theory with the force
computed using CEDIP, Tersoff and another DFT functional, PBE~\cite{perdew_1996}, as well
as the GAP from Ref.~\onlinecite{deringer_2017}, which was used in Ref.~\onlinecite{detomas_2019}
to generate the structure. The estimates above the GAP line indicate worse results
than GAP, whereas those estimates below are more accurate than GAP. (c) ICOOP
bonding/anti-bonding analysis for the 5 neighbors of all 5-c atoms, as
a function of bond length. ICOOP~$=0$ indicates the transition between bonding (ICOOP~$> 0$)
and antibonding (ICOOP~$< 0$). (d) ICOHP bond strength analysis for the 5 neighbors
of all 5-c atoms, as a function of bond length.}
\label{21}
\end{figure}

To gain further insight into the nature of these 5-c environments, we carried
out complementary analyses of geometry and electronic structure. On the one hand,
we estimated the force acting on the 5-c atom, predicted
by LDA-DFT (the reference in the a-C GAP), as a proxy for the stability
of these 5-c complexes. On the other hand, we quantify the chemical bonding nature using
crystal orbital overlap population (COOP)~\cite{hughbanks_1983} 
and crystal orbital Hamilton population (COHP)~\cite{dronskowski_1993} analyses, 
based on a local-orbital projection scheme as implemented in 
LOBSTER \cite{deringer_2011, maintz_2013, Nelson2020}.
In brief, a self-consistent electronic-structure computation in the projector-augmented wave
(PAW) framework~\cite{bloechl_1994} is carried out, here using VASP~\cite{kresse_1996, kresse_1999}.
The self-consistent electronic wave function is then projected onto an auxiliary, atom-centered
basis of 2$s$ and 2$p$ orbitals (following ideas proposed in Ref.~\onlinecite{SanchezPortal1995}), 
and the availability of local information allows the reconstruction of energy- and orbital-resolved
chemical-bonding indicators~\cite{deringer_2011, maintz_2013, Nelson2020}. 
The energy integration of 
COOP($E$) up to the Fermi level yields a measure for the
electron population associated with a given bond (positive values indicating stabilization),
whereas the integration of COHP($E$) gives an energy value (negative values indicating
stabilization) \cite{dronskowski_1993}. 
The projection onto a local basis makes it possible to analyze the output of
large-scale PAW-based DFT simulations of structurally complex materials~\cite{maintz_2013}, 
including studies of structure and bonding in the amorphous state~\cite{Deringer2014}.

We obtained a GAP-quenched a-C structure containing 25 such 5-c environments, out of a total
of 4096 C atoms (0.6\%), from the authors of Ref.~\onlinecite{detomas_2019}, who brought the issue of
5-fold coordinated C atoms in GAP simulations to our attention. Since computing energy
and forces for this structure at the DFT level is impractical, due to its large size,
we used a carving technique which involves passivation with H atoms to heal artificially
introduced dangling bonds~\cite{caro_2014}. The spherical clusters obtained in this
way [\fig{21}~(a)], carved
within a sphere of radius 7~{\AA} centered at the 5-fold atom, contained
an average of 239 C atoms and 145 H atoms. Comparing the force acting on the 5-fold atom from
LDA DFT to the GAP prediction (which is zero, since the structure is predicted by GAP to be
at equilibrium) gives an idea of the actual stability of the structure. As seen in \fig{21}~(b),
the errors range from 0 up to 7~eV/{\AA}, with most errors of the order of 2~eV/{\AA}. This analysis
indicates that the structures predicted by GAP are not fully stable in reality, but are not
totally unreasonable. In fact, a small fraction of these structures remained 5-fold coordinated
even after carrying out a structural relaxation at the DFT level of the innermost atoms in the
carved structure (3~{\AA} radius, with all other atoms between 3 and 7~{\AA} fixed).
The calculated root mean squared
displacement (RMSD) for these atoms (from the GAP-predicted to the DFT-relaxed structures),
averaged over the 25 5-c complexes, was only 0.10~{\AA}.

We further computed these
forces using the CEDIP and Tersoff force fields, observing much larger errors for those. This is
expected, since by construction traditional force fields introduce explicit biases regarding
coordination (e.g., that C should be 2-, 3- or 4-fold coordinated, but not 5-fold coordinated),
and disproportionately penalize structures which one does not expect according to classical
chemical rules. Unsurprisingly, differences between force fields are much larger than between
DFT functionals [\fig{21}~(b)].

The final test to elucidate the nature of these 5-fold complexes is the bond strength analysis
shown in Figs.~\ref{21}~(c) and (d). The results follow the expected bond-length--bond-strength
correlation (``shorter is stronger''), both for the overlap-based (ICOOP) and the Hamilton-matrix-based
(ICOHP) indicators. The trends for the first four neighbor contacts resemble those observed in a
comprehensive study of crystalline carbon allotropes~\cite{goerne_2019}, with seemingly slightly
weaker bonds (smaller ICOHP magnitude) in the amorphous than in the crystalline phases, 
not unexpectedly so.
In contrast, the analysis in \fig{21}~(c) suggests than the 5th neighbor of a given carbon atom
leads neither to substantial stabilizing nor to de-stabilizing orbital overlap (ICOOP $\approx 0$),
and concomitantly the associated contribution to the single-particle band-structure energy (a proxy
for the ``bond strength'', measured by ICOHP) is only a fraction of that of conventional
carbon--carbon bonds~\cite{goerne_2019}. These results might be taken to support the designation
of the corresponding atoms as ``4+1 coordinated'', i.e., with four strong bonds in a distorted
$sp^{3}$ configuration, but a fifth, much more weakly bonded atom still coming closer than
1.9~\AA{}. Additional, energy-resolved COHP results for the individual types of bonds are
provided in the SM.

\subsection{Computing the elastic properties}\label{18}

\begin{figure}[t]
\includegraphics[width=7.5cm]{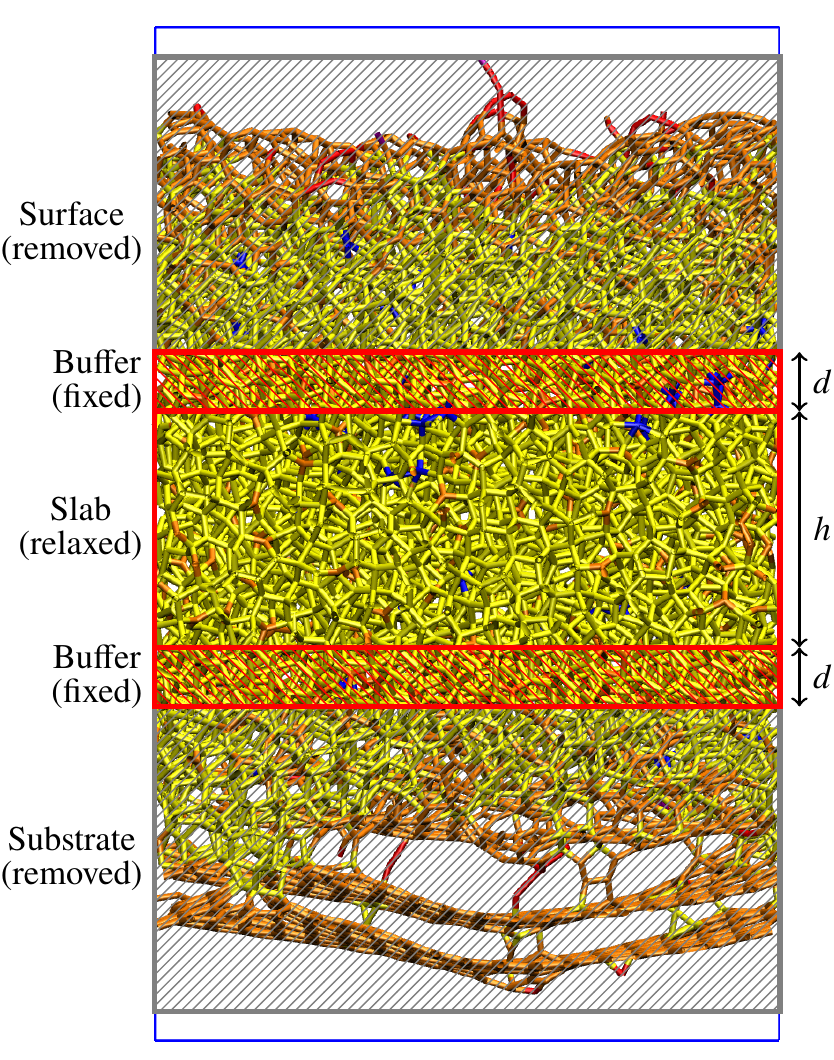}
\caption{Procedure to isolate the bulk-like portion of the atoms in the slab for the computation
of elastic properties. See text for details.}
\label{04}
\end{figure}

\begin{figure*}[t]
\includegraphics{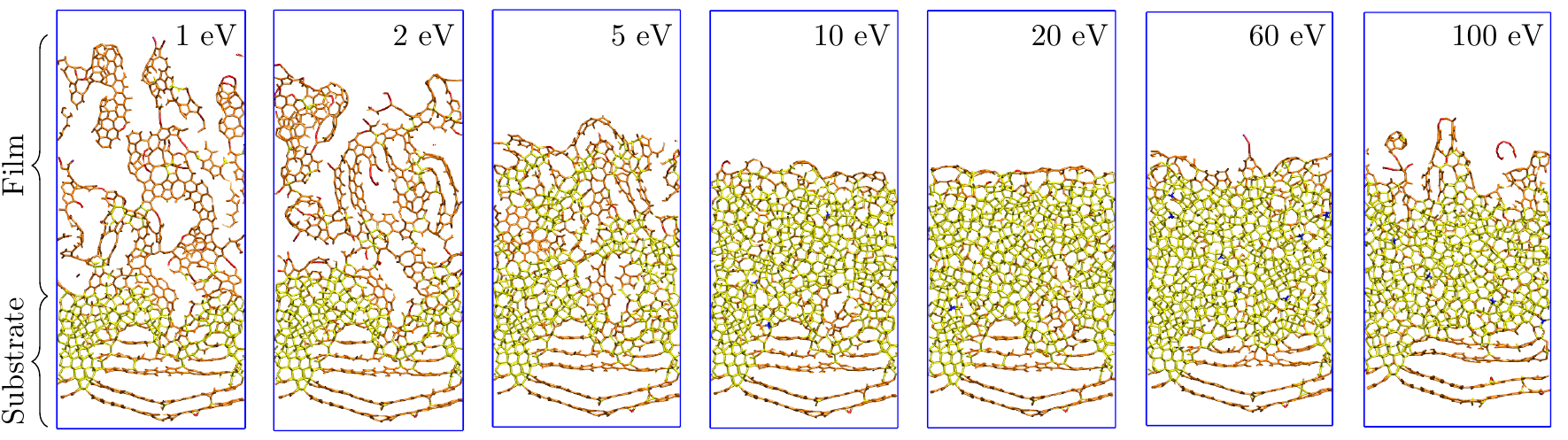}
\caption{Amorphous carbon films grown by ML-driven deposition simulations, varying the energy of
the impacting ions over a wide range from 1 eV to 100 eV. Structures are shown as cross sections,
corresponding to 4~{\AA} thick slices of the grown films (only), to emphasize the graphitic-like
features of the low-density films. 
Red, orange and yellow atoms represent $sp$, $sp^2$
and $sp^3$ hybridization, respectively. Other colors represent different coordination
defects (5-fold coordination in blue and 1-fold coordination in purple).}
\label{13}
\end{figure*}

To compute the elastic properties of the films, we first carried out a quenching from 300~K
to close to 0~K. After this, a geometry optimization followed. All elastic properties were
computed for these quenched structures at zero temperature. Since our a-C samples are grown
as films, computing the \textit{bulk} elastic properties is not straightforward. This is
because, under periodic boundary conditions, one needs to devise a strain transformation on
the supercell which discards the contribution to the elastic constants arising from the
surface and substrate. Carelessly taking the atoms in the central region of the film and
calculating elastic properties with them results in problems associated with (i) the loss
of periodicity  along the growth direction and (ii) surface reconstruction effects. Here,
instead, we introduce and take the following approach.

We select a group of atoms in the central part of the film where its properties are converged
and bulk-like (e.g., the $sp^3$ content does not change). The atoms within the center, in a
region of thickness $h$, are allowed to fully relax.  Atoms at the top and bottom of this
group, within a thickness $d$, are frozen, except for a possible strain transformation
(without further optimization, this is known as the ``clamped-ion'' approximation). All other
atoms, even further away from the central layer, are removed from the simulation box.
Figure~\ref{04} provides a schematic view. This procedure is repeated for different values of
$h$ and the evolution of the system's energy is monitored. For large enough supercells (that
is, large enough numbers of atoms), both the  energy \textit{density} and the surface energies
should be conserved quantities. By fitting the computed energies to the following equation of
state, we can compute the strain dependence of the energy:
\begin{align}
E^\text{slab}(\epsilon) = \frac{\partial E^\text{slab} (\epsilon)}{\partial h} h +
E^\text{buffer}(\epsilon; d),
\label{07}
\end{align}
where $\epsilon$ denotes the full strain tensor or, rather, the vector containing the 6 independent
Voigt components of the strain tensor, $\epsilon_i$. $E^\text{buffer}(\epsilon, d)$ is an
energy term related to the top and bottom surfaces and interfaces (broken bonds, frozen atoms, etc.)
which does not change with $h$.

The advantage of this expression is that, to compute bulk elastic properties, only
$\frac{\partial E}{\partial h}|_\epsilon$ is required, because it fully characterizes the elastic
response of the bulk. In other words, we have
\begin{align}
\lim\limits_{h \rightarrow \infty} \frac{\partial E^\text{slab}}{\partial h}|_\epsilon =
\lim\limits_{h \rightarrow \infty} \frac{\partial E^\text{bulk}}{\partial h}|_\epsilon \, .
\end{align}
Furthermore, since the GAP relies on cutoff distances to define atomic interactions, choosing
$d$ to be at least as large as the cutoff (here, 3.7~{\AA}) means that the interactions within
$h$ are preserved in the carved slab, as compared to the original film structure. The elastic
constants, $C_{ij}$, for the bulk-like region in the film center can then be computed as
\begin{align}
C_{ij} & = \frac{1}{V_0} \frac{\partial^2 E^\text{bulk}}{\partial \epsilon_i \partial \epsilon_j}
\nonumber \\
& =  \lim\limits_{h \rightarrow \infty} \frac{1}{A_0} \left. \left(
\frac{\partial^2}{\partial \epsilon_i \partial \epsilon_j}
\frac{\partial E^\text{slab} (\epsilon)}{\partial h} \right) \right|_{\epsilon = 0},
\label{05}
\end{align}
where $V_0$ and $A_0$ are the volume and cross-sectional area \textit{at equilibrium}. Here
one should emphasize what equilibrium means, since a-C films are under compressive biaxial
stress. We define equilibrium as the geometry of the film \textit{as grown}, that is, under
compressive stress. At this geometry, the system is not at the minimum of the bulk
energy-versus-strain curve (otherwise the stress would be zero). These elastic constants
should be directly comparable with experiment. At some other strain $\epsilon$, the effective
elastic constants are given by
\begin{align}
C_{ij} (\epsilon) =  \lim\limits_{h \rightarrow \infty} \frac{1}{A(\epsilon)} \left.
\left( \frac{\partial^2}{\partial
\epsilon_i \partial \epsilon_j} \frac{\partial E^\text{slab} (\epsilon)}{\partial h}
\right) \right|_{\epsilon}.
\end{align}
The stress is given by the first derivative of the energy at $\epsilon$:
\begin{align}
\sigma_i (\epsilon) =  \lim\limits_{h \rightarrow \infty} \frac{1}{A(\epsilon)}  \left.
\left( \frac{\partial}{\partial
\epsilon_i} \frac{\partial E^\text{slab} (\epsilon)}{\partial h} \right) \right|_{\epsilon}.
\end{align}

At this point we need to make a remark of practical importance. The energy changes much more
quickly by adding more atoms at fixed strain (increasing $h$) than by applying strain at
fixed number of atoms (fixed $h$); therefore, fitting the data to \eq{07} directly turns out
to be impractical. Instead, we choose to change the order of partial derivatives, so that
the quantity (numerically) evaluated is the evolution of $C_{ij}$ and $\sigma_i$ with $h$,
which are much smoother than the evolution of $E$ with $h$:
\begin{align}
C_{ij} (\epsilon) =  \lim\limits_{h \rightarrow \infty} \frac{1}{A(\epsilon)}
\frac{\partial}{\partial h} \left. \left(
\frac{\partial^2 E^\text{slab} (\epsilon)}{\partial \epsilon_i \partial \epsilon_j} \right)
\right|_{\epsilon}
\label{25}
\end{align}
and
\begin{align}
\sigma_i (\epsilon) =  \lim\limits_{h \rightarrow \infty} \frac{1}{A(\epsilon)}
\frac{\partial}{\partial h} \left. \left(
\frac{\partial  E^\text{slab} (\epsilon)}{\partial \epsilon_i}  \right) \right|_{\epsilon},
\end{align}
respectively, where the quantities in brackets are evaluated first. We have assumed that \eq{07} holds;
that is, we can write:
\begin{align}
& C_{ij}^\text{bulk} (\epsilon) = \lim\limits_{h \rightarrow \infty} \frac{1}{A(\epsilon) h}
\frac{\partial^2 E^\text{bulk} (\epsilon; h)}{\partial \epsilon_i \partial \epsilon_j}
\nonumber \\
& = \lim\limits_{h \rightarrow \infty} \frac{1}{A(\epsilon) h} \frac{\partial^2
\left(E^\text{slab+buffer}(\epsilon; h,d) - E^\text{buffer} (\epsilon; d) \right)}{\partial
\epsilon_i \partial \epsilon_j}
\nonumber \\
& = \lim\limits_{h \rightarrow \infty} \frac{1}{A(\epsilon) h} \left(\alpha_{ij} (\epsilon;
\mathcal{H}) h + \beta_{ij} (\epsilon; d, \mathcal{H}) - \beta_{ij} (\epsilon; d, \mathcal{H})
\right)
\nonumber \\
& = \lim\limits_{h \rightarrow \infty} \frac{1}{A(\epsilon)} \alpha_{ij} (\epsilon; \mathcal{H}).
\end{align}
The $\alpha_{ij}$ and $\beta_{ij}$ are simply the coefficients resulting from a linear fit of
$\partial^2 E^\text{slab+buffer} (\epsilon; d, \mathcal{H}) / \partial \epsilon_i \partial
\epsilon_j$ versus $h$, for a fixed value of $d$. The second derivatives of the energy,
\eq{25}, are themselves obtained from a second-order polynomial fit of the energy
on a 25-point ($5 \times 5$) 2D mesh of the strain components, at 0.2\% strain
increments ($-0.4$\% to +0.4\%). Given the symmetry of the films (further discussed in the
appendix), we choose strain branches corresponding to $[\epsilon_1 = \epsilon_2 \neq \epsilon_3]$
and $[\epsilon_1 \neq \epsilon_2; \epsilon_3 = 0]$. Finally, note that $\alpha_{ij}$
depends on the fitting domain $\mathcal{H} \equiv [h_\text{min}, h_\text{max} ]$. This
dependence is weak if a suitable domain is chosen (i.e., $h_\text{min}$ is large enough).

\section{Results and discussion}

\subsection{Simulated carbon films throughout the entire density range}

\begin{figure}[t]
\includegraphics{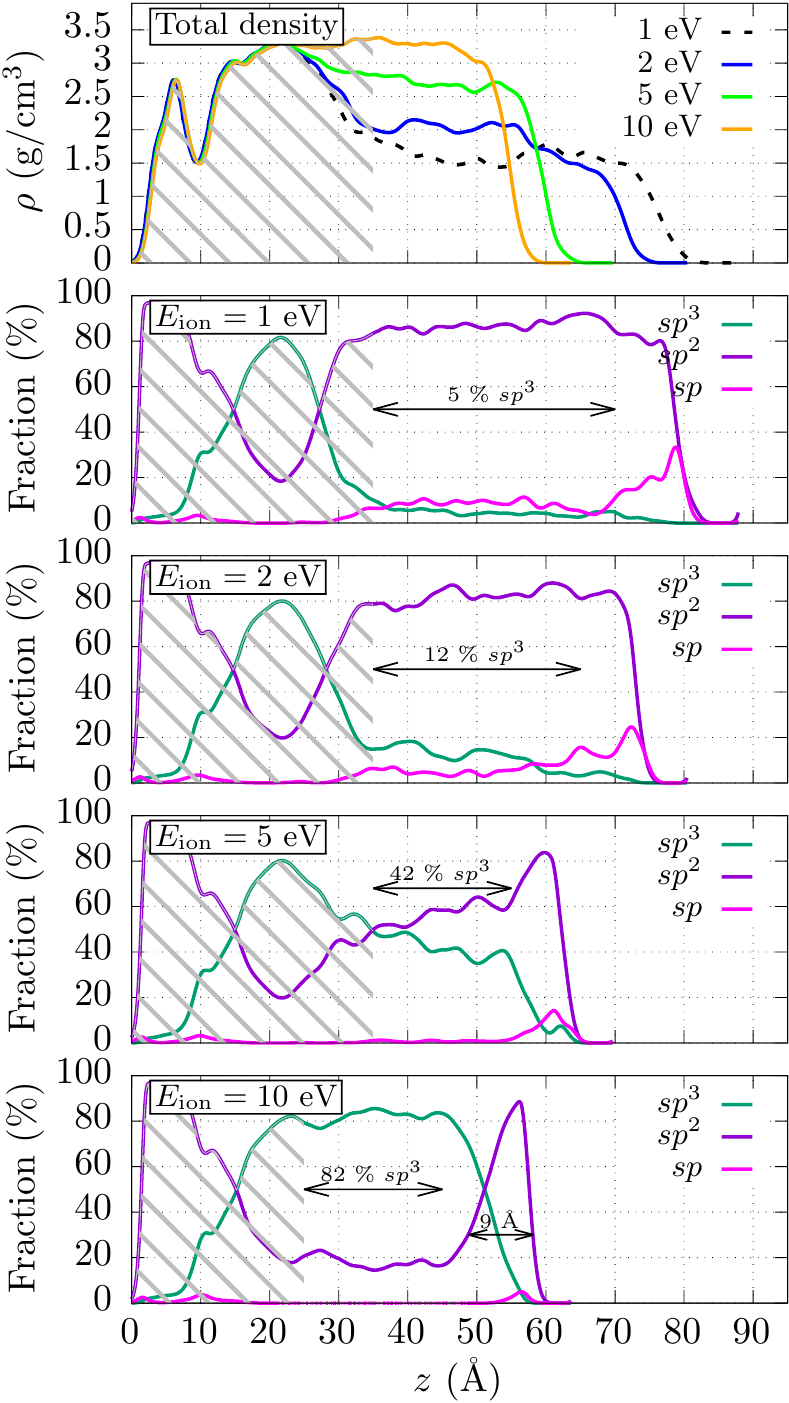}
\caption{Total mass density ({\em top}) and coordination profiles ({\em bottom}) for the films
grown at low deposition energies, plotted in the same way as in Fig.\ \ref{06} (that is, with
the horizontal axis following a slice through the slab, and the substrate region indicated by
shading). Very low-density a-C films, well below the density of graphite, are created
at 1--2~eV. At 5 eV, a coexistence of $sp^{2}$ (purple line) and $sp^{3}$ (green line)
environments is observed in the center of the slab; at 10~eV, the film is already very rich in
$sp^{3}$-bonded carbon atoms (albeit not having reached the $\approx 90\%$ of ta-C), and a
distinct $sp^{2}$-rich top layer appears. Coordination is based on counting the number of
neighbor atoms within a cutoff sphere of radius of 1.9~{\AA}. The shaded areas indicate the
portion of the films corresponding to the initial substrate (the same substrate is used for all
deposition energies, see text).}
\label{20}
\end{figure}

Our simulations were performed over a wide range of deposition energies under otherwise
similar conditions. This allows us to carry out a comprehensive characterization of all
possible types of deposited a-C, from very low density a-C (1.5~g/cm$^3$ at 1~eV, 
$\sim 65$~\% of the density of graphite) all the way up to ultra-high density ta-C 
(3.4~g/cm$^3$ at 20~eV and beyond, $\sim 96$~\% of the density of diamond); it also allows
us to assess the effect of ion energy on the surface structure in a systematic fashion.
The main object of study is a series of seven distinct slab models at deposition energies of
1, 2, 5, 10, 20, 60 and 100~eV (Fig.~\ref{13}). These are the results thoroughly discussed throughout
this paper: an additional four simulations at 3, 4, 6.5 and 8~eV were conducted, which
are reported in the SM in the interest of clarity. 
The impact energy is a quantity which can
be directly controlled in experiment, and it therefore constitutes a possible avenue to
``design'' carbon materials. 

In Fig.~\ref{13} we show cross-sectional slices (4~{\AA} thick) through the films. 
We can clearly observe the morphological and coordination changes
that take place in a-C as the deposition energy increases. At low energy/density
(1 and 2~eV, 1.5~g/cm$^3$ and 2~g/cm$^3$, respectively), the
a-C films are composed of loosely connected tubular (nanotube-like) $sp^2$ structures,
qualitatively resembling existing models of ``glassy'' carbon (Ref.~\onlinecite{harris_2005}
and references therein), as well as the result of earlier graphitization simulations starting
from bulk a-C~\cite{powles_2009, palmer_2010, detomas_2017, detomas_2018}.
As the deposition energy and density increase, the material takes the form of tightly
embedded $sp^2$-rich regions a few {\AA} across in an $sp^3$-rich matrix
(5~eV, 2.6~g/cm$^3$). We note the conceptual similarity of such coexisting regions
to the results of GAP-driven simulations reported by de Tomas et al.~\cite{detomas_2019},
and the experimental observation of nanoscale-ordered $sp^2$/$sp^3$ composite materials
by transmission electron microscopy~\cite{hu_2017, nemeth_2020}. 

These results are presented more quantitatively in the mass density
and atomic coordination profiles; they had been given in our earlier study~\cite{caro_2018}
and in Fig.~\ref{06} for high densities, and we now show the equivalent plots for 
low-density structures in Fig.~\ref{20}. We reiterate that the bottom region of the simulation
systems is pre-determined by the substrate, and these regions are therefore shaded in Fig.~\ref{20}.
The density change at (very) low impact energies is directly mirrored by a larger spatial
extent of these slabs. It is noteworthy that at 1 and 2 eV, a non-negligible
amount of $sp$ atoms persists throughout the low-density part of the slab, whereas this
coordination mode is only seen in the surface layer ($z > 50$ \AA{}) for the 5 eV
deposition, and almost not at all at 10 eV. Another noteworthy observation is that for
the outcome of the simulation performed at 5 eV, 
$sp^{2}$ and $sp^{3}$ atoms coexist in similar amounts.

\begin{figure*}[t]
\includegraphics{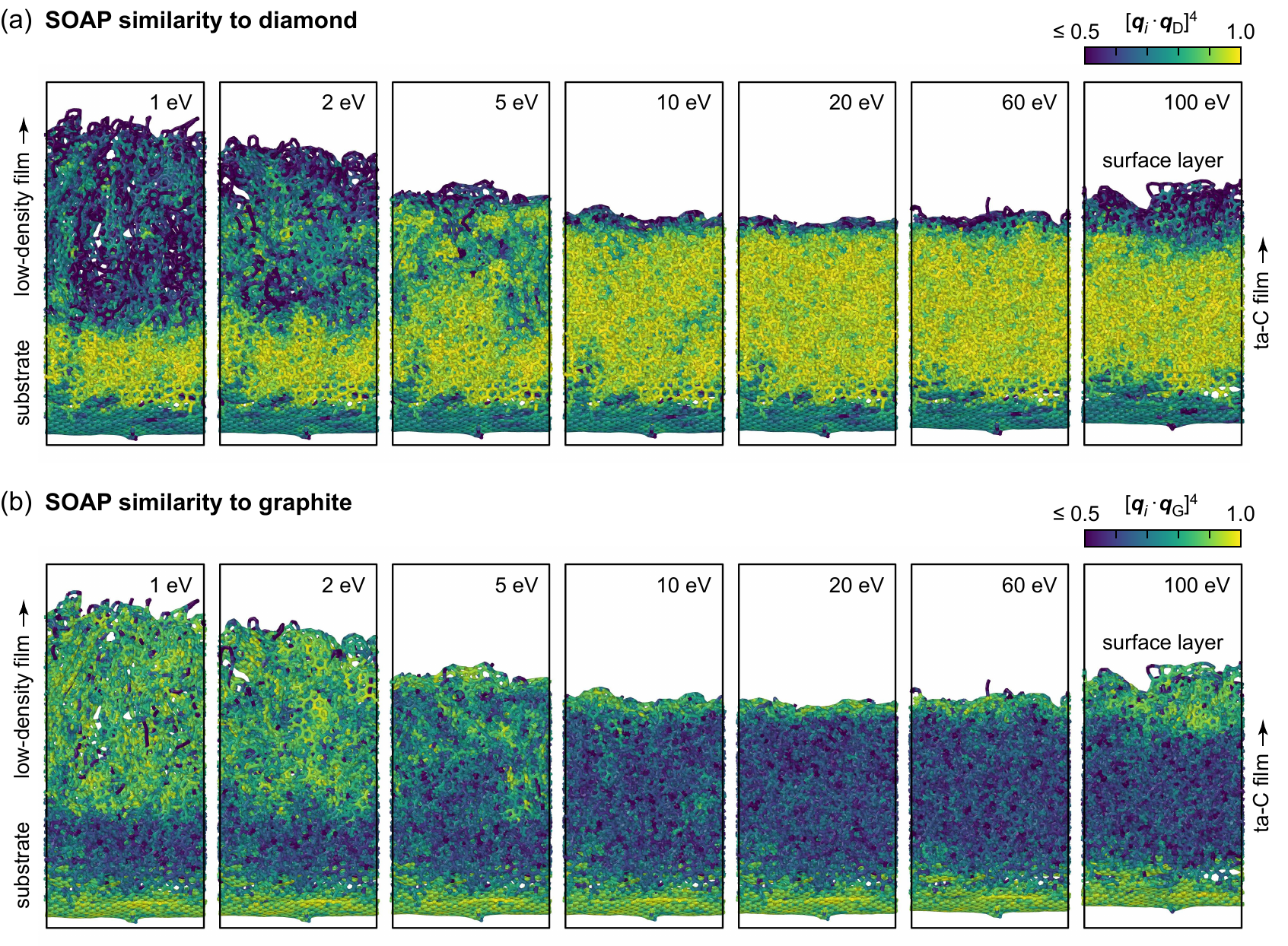}
\caption{Structural insight into the simulated carbon films by atom-resolved SOAP
similarity analysis~\cite{deringer_2018}. We compute the SOAP power-spectrum vector,
${\bf q}_{i}$, for every $i$-th atom in a given system, and evaluate the dot product
of this vector with its equivalent for ideal diamond and graphite, respectively (raised
to the power of 4 to enable a better distinction between environments). This yields a
similarity value between 0 (entirely unlike the reference crystal) and 1
(identical within the cutoff radius), which is indicated by color coding. We use a SOAP
cutoff radius of 3.7~\AA{}, the same as in the fitting of the GAP model~\cite{deringer_2017},
and a highly converged number of basis functions ($n_{\rm max} = l_{\rm max} = 16$). Note
that the amount of vacuum in some of the supercells has been increased to ease
visualization.}
\label{soap-colourcoded}
\end{figure*}

It is also interesting to quantify the similarity
to the ideal diamond and graphite structures using SOAP~\cite{bartok_2013}, which we have
previously demonstrated for small samples of ta-C~\cite{deringer_2018} and amorphous
silicon \cite{bernstein_2019}. With much larger simulation systems now available, we may
assess the ``diamond-likeness'' and ``graphite-likeness'' of our a-C systems in a systematic
fashion, including realistic estimates of the effect of film thickness. Color-coded plots,
akin to \fig{13} but now with the additional structural information provided by SOAP,
are shown in \fig{soap-colourcoded}. We note that this type of analysis yields a continuous scale
for quantifying the short- and medium-ranged structural environments of individual atoms, 
which is expected to be more nuanced than the established convention of counting nearest
neighbors and assigning ``$sp^{2}$'' and ``$sp^{3}$'' nature based on that (see also
Ref.~\onlinecite{caro_2018c} in this context).

We recall that all simulations start from the same substrate, viz.\ a thin ta-C film (obtained
by deposition on a diamond-type surface, which is fully disordered in the process, and forms
graphite-like sheets at the bottom of the slab). Therefore, the substrate is clearly made up
of a diamond-like region [yellow in \fig{soap-colourcoded} (a)] and terminated by a thin
graphite-like region at the bottom [yellow in \fig{soap-colourcoded} (b)]. It is above this
surface that we observe very distinct structural properties and trends as the deposition energy
is being varied.

The low-energy simulations (1--2 eV) lead to a low-density film (as already apparent from
the density profiles in \fig{20}), which is very dissimilar to diamond but locally
resembles graphite. The film at 5 eV is perhaps the most interesting, because it shows clearly
distinct regions of diamond- or graphite-likeness [that is, complementary regions
are ``lighting up'' in Figs.~\ref{soap-colourcoded} (a) and \ref{soap-colourcoded} (b),
respectively]. Between 
10 and 60 eV, relatively uniformly diamond-like films are obtained, with concomitant very low
similarity to graphite. At 100 eV, we observe the formation of a thicker surface layer
(cf. density profiles in \fig{06}), and this one is again similar to graphite. 

\begin{figure}[t]
\includegraphics{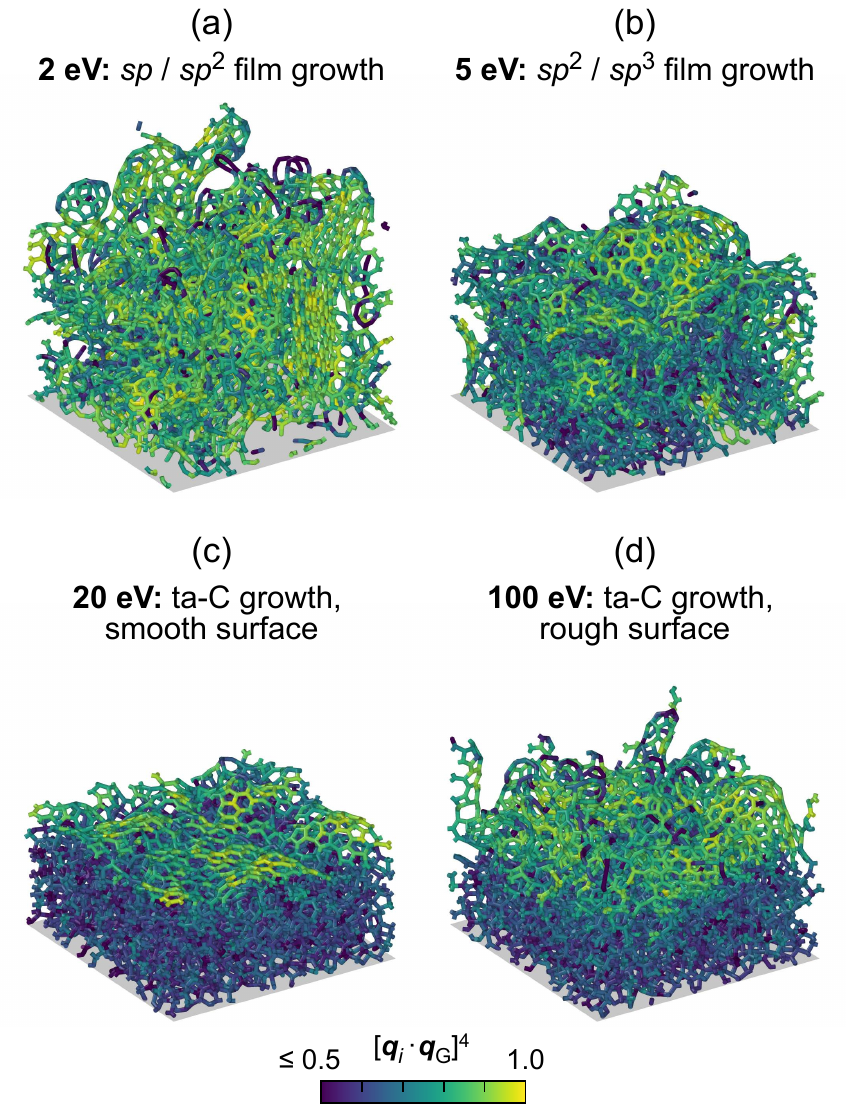}
\caption{Details of the surface regions of our simulated a-C systems, indicating four distinctly
different types of film obtained at various ion energies; color-coding shows the SOAP similarity
to graphite as in \fig{soap-colourcoded}. The bottom 35 \AA{} of each simulation cell
(containing the substrate) has been removed to ease visualization.}
\label{surface_gra_details_compr}
\end{figure}

We provide more detailed close-ups of the film structures in \fig{surface_gra_details_compr},
which allows us to identify distinct qualitative types of a-C films as dependent on the incident ion 
energy. Movies showing the growth of these films are provided online~\cite{caro_2017d} and
in the SM of Ref.~\onlinecite{caro_2018}. The atomic structures resulting from the deposition
simulations are also provided online in extended XYZ format~\cite{caro_2020}, in the hope that
they may enable further work in the future.
Indeed, libraries of (small-scale) GAP-generated carbon structures have begun to be successfully 
used as starting points for simulation studies by others \cite{Lahrar2020, Wang2020a}.

\subsection{Growth mechanisms at low and high density}

\begin{figure}[t]
\includegraphics[width=7.5cm]{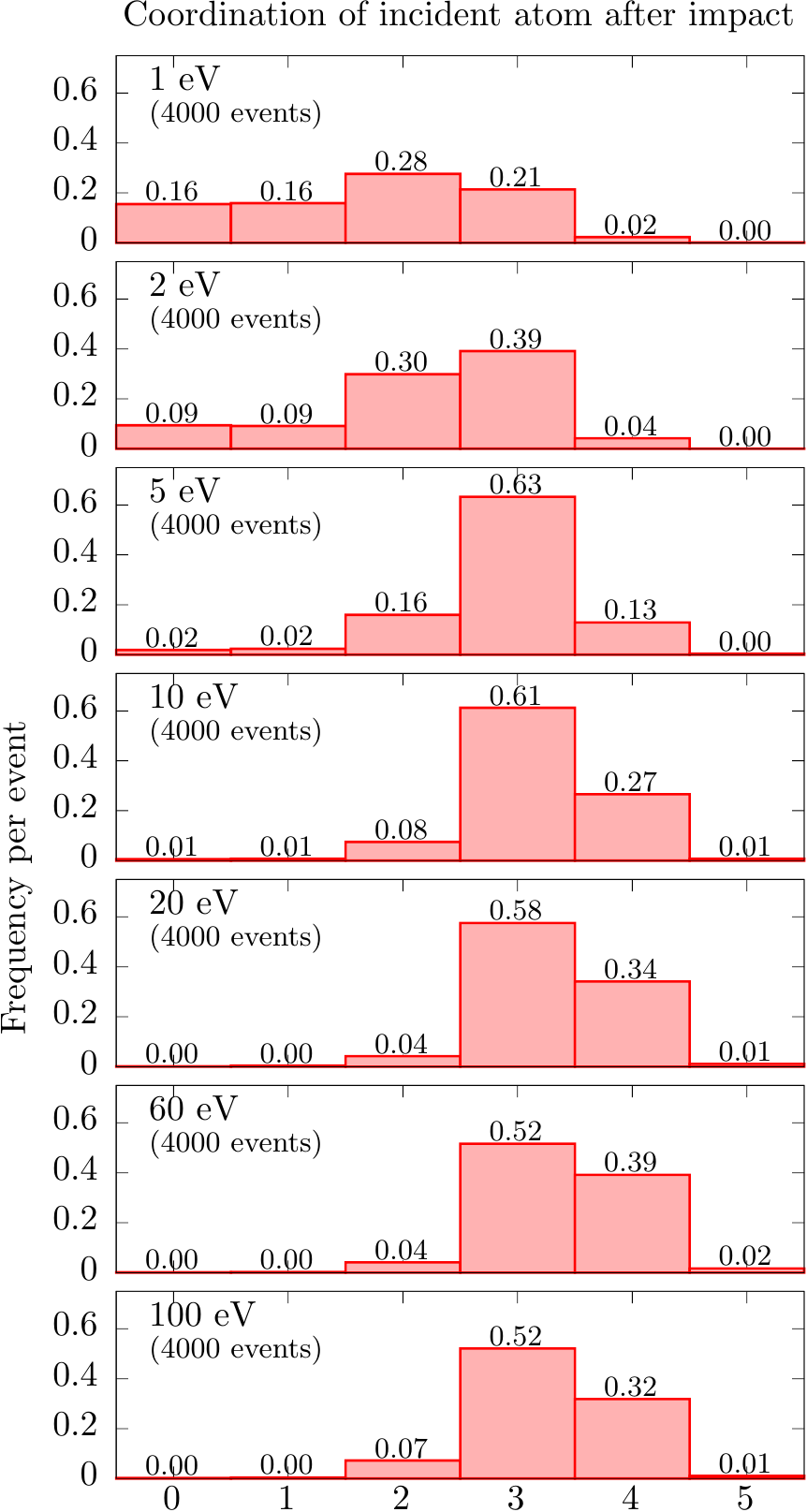}
\caption{Coordination numbers of incident atoms after impacting the surface. Statistics for 4,000
events in each individual simulation are given. At the lowest energy, 1 eV, a substantial number
of atoms ($\approx$ 16\%) exhibit zero bonding partners after impact: these atoms dissipate from
the surface and have therefore been removed from the simulation cell, repeating the simulation with
a new impact event. Hence, it should be noted that the connectivity in the final films is different
from the numbers collected here.}
\label{14}
\end{figure}

\begin{figure}[t]
\includegraphics{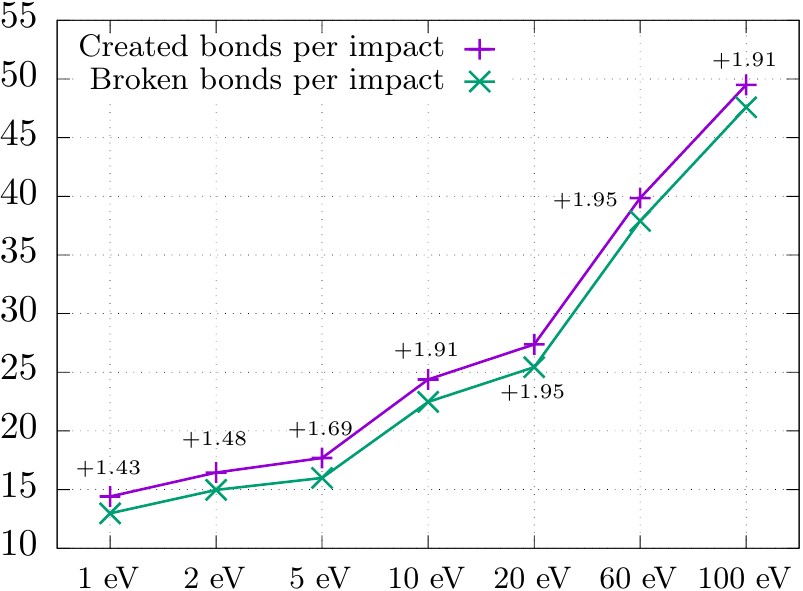}
\caption{The balance between the number of bonds that are broken and created per impact event
(averaged over the last 4000 impacts at each given energy). The absolute numbers are plotted on
the vertical axis, and they range from about 15 bonds at low energy to about 50 bonds {\em per
impact event} at high energy, emphasizing the many complex structural transformations that take
place throughout the cell, especially at high energy (where ta-C grows). It is then instructive to
inspect the {\em difference} between the absolute numbers of bonds created and broken, and this
difference is given by labels on the individual data points. These differences are roughly consistent
with the average creation of 3/2 bonds (i.e., a new threefold-coordinated $sp^{2}$ environment) at
low energy, and of 4/2 bonds (fourfold-coordinated, $sp^{3}$) at high energy.
A background number of rebonding events due to thermal fluctuations, which is proportional to the
number of atoms in the film, has been subtracted (see text for details).}
\label{15}
\end{figure}

\begin{figure}[t]
\includegraphics{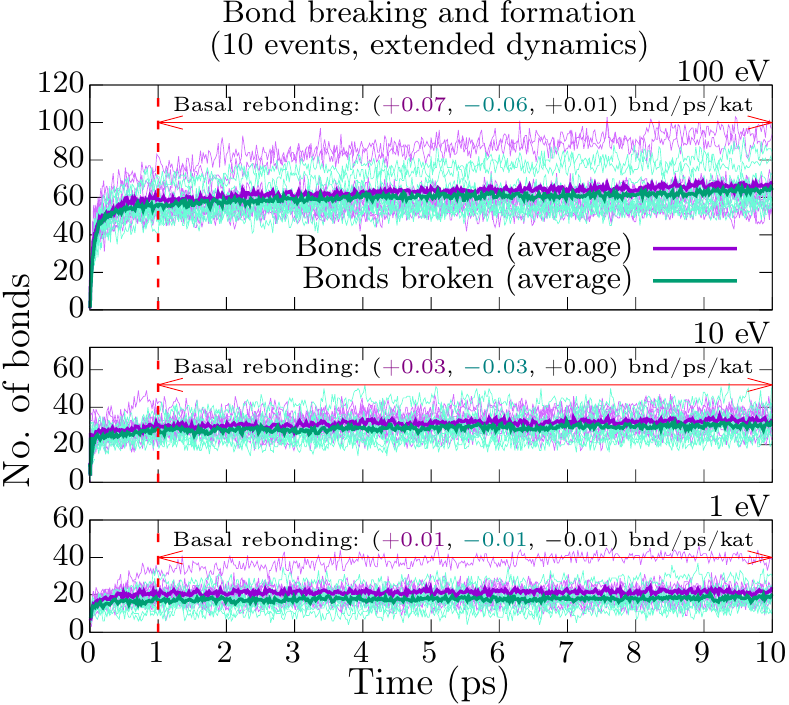}
\caption{Rebonding computed for a sample of 10 impacts at three different deposition
energies (1, 10 and 100~eV), studied over an extended MD equilibration period (10~ps versus
1~ps for all other data). Beyond the 1~ps mark employed for equilibration in our non-extended
simulations, the number of rebonding events is very small, less than 0.1 rebonding
events per ps and per thousand atoms, ``kat''. Bond formation, bond breaking and net effect
during the post-equilibration period are indicated as ``basal rebonding'' with purple, green and black
numbers, respectively, with the net effect being almost negligible.
This means that the films are relatively stable and remain so
after the initial impact and rearrangement events have taken place. Purple lines indicate created
bonds and green lines indicate broken bonds. Individual data are shown
with light thin curves and average (over 10 events) data are shown with darker thick lines.}
\label{26}
\end{figure}

\begin{figure*}[t]
\includegraphics{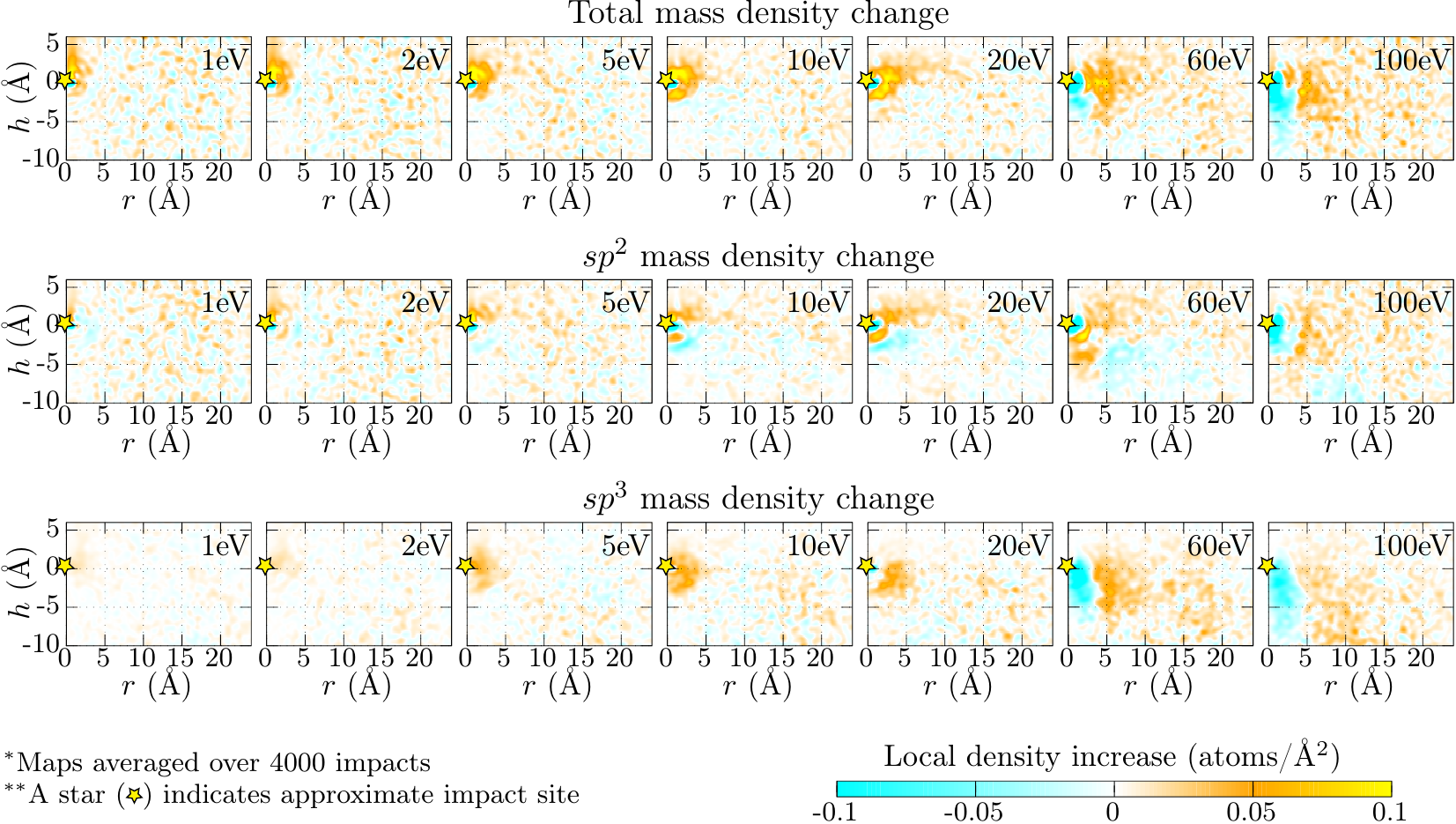}
\caption{Mass density change maps, as a function of deposition energy, computed using \eq{17} (see
discussion of the equation for further details). The mass density increase/decrease is also split
into partial $sp^2$ and $sp^3$ mass densities. One can observe that the deposition mechanism changes
as the energy increases. At low energy, incoming atoms are deposited near the impact site. At high
energy, mass density is locally depleted around and below the impact site and $sp^3$ carbon is formed
laterally and away from it. At high energy, this deposition mechanism is known as ``peening'', and
is discussed in Refs.~\cite{robertson_2002,marks_2005,caro_2018}. The line of
impact ($r=0$) corresponds to incident atom's initial $xy$ coordinates, whereas
the height of impact ($h=0$) corresponds to the $z$ coordinate of the first
atom it encounters within an impact cylinder of radius 1~{\AA}.}
\label{16}
\end{figure*}

\begin{figure}[t]
\includegraphics{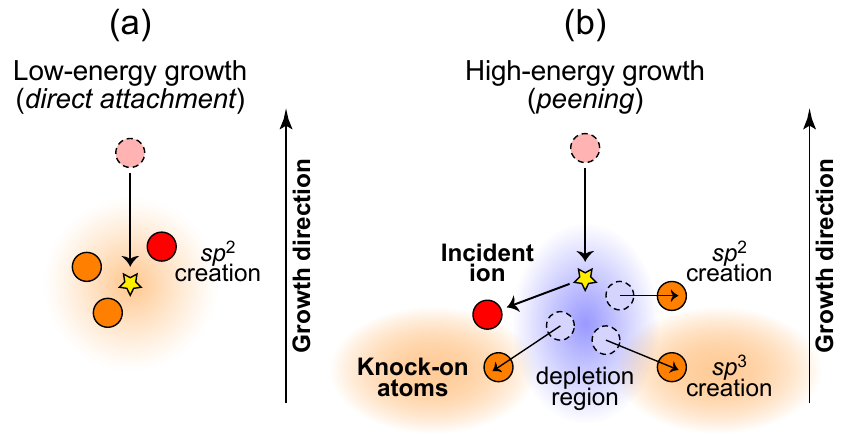}
\caption{Growth mechanism at low and high impact energy as deduced from the present simulations.
(a) Proposed growth mechanism at low density. 
(b) Illustration of the ``peening'' mechanism (increased atomic coordination
takes place laterally and away from the impact site due to pressure waves)
which according to our simulations is consistent with the growth of high-density
ta-C films. Adapted from Ref.~\onlinecite{caro_2018}.}
\label{growth_mechanism_sketches}
\end{figure}

Amorphous carbons exist within a wide range of experimental densities, which correlate
strongly with the fraction of $sp^3$-bonded carbon atoms, that is, carbon atoms with four
neighbors. This is the same bonding configuration exhibited in crystalline diamond.
Therefore, the densest a-C samples, referred to as ta-C or diamond-like carbon (DLC),
also show the highest $sp^3$ fractions. Superhard ta-C can contain up to 90\% of $sp^3$
atoms and reach densities and elastic properties very close to those of diamond. The growth
mechanism leading to such high $sp^3$ contents was poorly understood until very recently.
Using the same methodology as here, we were able to elucidate the ta-C growth mechanism
in a previous study. Contrary to the assumption prevalent in parts of the literature, we
showed~\cite{caro_2018} that ta-C grows \textit{predominantly} by ``peening''~\cite{marks_2005}
[\fig{growth_mechanism_sketches} (b)],
rather than ``subplantation''~\cite{robertson_2002}. 
In short, the subplantation mechanism
assumes that the increased atomic coordination in high-density a-C (high $sp^3$ content)
is due to atomic packing taking place locally at the site of impact of the deposited atoms.
In contrast, our previous simulations showed that locally (at the impact site) the density
of the film decreases after each deposition event, creating an increased likelihood of
$sp^2$ formation within a ``depletion region'' about 5~{\AA} wide,
and local \textit{destruction} of preexisting $sp^3$ sites in this region. The
locally-displaced atoms (incident and knock-on atoms) lead to a pressure wave outward from
the impact site and the subsequent packing of C atoms laterally and away from this
site. This pressure-led
packing is the predominant mechanism responsible for increased coordination in the
subsurface region
of the film. The peening mechanism was originally proposed by Marks based on CEDIP
deposition results~\cite{marks_2005}. However, the limited quantitative agreement of CEDIP
with experimental data for the high $sp^3$ fractions in these films
prevented the widespread adoption of this model.
Our results with the more flexible and arguably accurate (yet computationally more expensive)
GAP potential suggest that, while not in full agreement in all
quantitative ($sp^3$ fractions) and qualitative (surface characteristics) details, Marks'
CEDIP results were actually reproducing the correct deposition physics~\cite{caro_2018,
marks_2005}.

Having solved the issue of understanding the growth mechanism of high-density samples, here
we focus on gaining insight to the microscopic picture of a-C formation throughout the full
spectrum of mass densities. This will allow us to establish the physical mechanism responsible
for the transition from the $sp^2$-rich regime to the $sp^3$-rich regime as the deposition
energy is increased. In the future, this knowledge may prove key in optimizing the synthesis
of nanoforms of carbon with targeted properties in mind.

Movies showing the growth of these films are given in Ref.~\onlinecite{caro_2017d}.
In \fig{13} we show the state of each of the simulated films at the end of the deposition.
These cross-sectional figures show the general features of the films and depict the
transition from highly graphitic-like films at low deposition energy to diamond-like films
at high deposition energy. Having access to the full range of deposition energies and
resulting mass densities grants us insight into the changes that drive the transition from
$sp^2$-rich to $sp^3$-rich a-C. We proposed two basic analysis tools to study the growth
mechanism in a-C \cite{caro_2018}: i) coordination distribution analysis for incident atoms after impact
and ii) $sp^2$- and $sp^3$-resolved mass density maps highlighting local changes in atomic
coordination before and after impact.

The first tool allows us to establish what happens to the incident atom after impact. There
are essentially two main possibilities: either the atom is incorporated to the growing film
or it bounces off the surface. In both cases, the final status of the incident atom can
be characterized in a simple manner by its coordination after impact, where 0 coordination
indicates that the atom bounced off. These numbers are summarized, for the different deposition
regimes studied, in \fig{14}. As can be seen from the figure, there is a significant proportion
of incident atoms that bounce off at low deposition energy. This is easy to understand in
intuitive terms since low-energy incident atoms may a) not have enough kinetic energy to
climb over free-energy barriers required to become bonded to the substrate or b) not be
able to travel far enough into the film so as to become trapped inside until the conditions
are favorable for them to become bonded to the a-C matrix. Hence, as many as 16\% of all
incident atoms bounce off at 1~eV deposition energies. This number goes below 1\% at 20~eV
and higher energies. For those atoms that get implanted in the growing film, \fig{14} lets us
visualize what happens to them. In particular, incident atoms are implanted predominantly
with approximately 2-fold ($sp$) average coordination in the very-low energy regime
(1~eV). As the deposition energy is increased, the proportion of atoms that are deposited
with 3-fold ($sp^2$) and 4-fold ($sp^3$) coordinations increases. However, it is important
to note that, above 1~eV, 3-fold coordination always dominates over 4-fold coordination
as the state of the incident atom after deposition. In fact, the fraction of 4-fold deposited
atoms increases from 2\% at 1~eV up to a maximum of 39\% at 60~eV, but then decreases again 
at higher energies. In other words, if a-C grew by subplantation as postulated in the
literature for the past 30 years~\cite{robertson_2002}, films with $sp^3$ fractions in
excess of approximately 40\% would not be possible. We know from experiment that the
maximum $sp^3$ fractions attainable under optimal growth conditions are in excess of 80\%,
and up to 90\% for superhard ta-C films~\cite{schultrich_1998}. This fact alone disproves
subplantation as the main mechanism responsible for ta-C growth~\cite{caro_2018}.

The average number of bonds broken and created upon each deposition event
is shown in \fig{15}. Looking in more detail at the process of bond creation and
annihilation affords us extended understanding of the delicate balance between the different
chemical reactions taking place during a-C growth. From the figure we see how in the region
relevant to ta-C growth up to circa 50 bonds are broken during each deposition event (i.e.,
between impact and the end of the subsequent equilibration), with an average net creation of
approximately 2 bonds per event. This further highlights the remarkable success of the GAP,
correctly predicting the $sp^3$ formation rate despite the fact that it is a small net
effect between bond creation and annihilation. Even at low deposition energies we can observe
a significant dynamical balance between the two processes. In view of these numbers, it is
perhaps less surprising that a highly sophisticated interatomic potential is needed to
simulate a-C growth, given the sheer complexity of the dynamical equilibrium between all the
chemical reactions that follow in cascade each deposition event. Note that we have removed
from the graph the effect of statistical thermal fluctuations on rebonding processes in the
films. This effect is easily subtracted from the data since it is linearly proportional
to the system size (this effect is below 1.5 broken/created bonds per 1000 atoms per ps).

Under experimental conditions, the rate of deposition is much lower than 1 atom per ps.
Unfortunately, due to the current computational limitations, we cannot afford to run equilibration
times comparable to experiment. To probe what would happen to the films under more realistic
conditions, we have repeated a series of 10 deposition events at 1, 10 and 100~eV, and let the
system equilibrate up to 10~ps (i.e., approximately one order of magnitude longer). The results,
shown in \fig{26}, indicate that the films remain rather stable after the initial impact event
followed by strong rearrangement of the local atomic environments. We take this as a sign
that the employed equilibration times are sufficient to capture the leading processes determining
the final atomic structure of the films.

The second tool at our disposal offers detailed insight into the actual growth
mechanism in a-C, and allows us to put the results from \fig{13} and \fig{14} into context. This
tool is based on the differences between pair correlation functions (PCF) (split into $sp^2$ and
$sp^3$ components) computed before and after impact~\cite{caro_2018}. In essence, we compute a
two-dimensional PCF where the first dimension is height from impact point $h$ and the second
dimension is radial distance from line of impact $r$. This PCF, $g(r,h)$,
is thus given in cylindrical
coordinates and is therefore adapted to the expected cylindrical symmetry of the film's
characteristics about the high-symmetry line corresponding to the incident trajectory of the
impacting atom. The difference between this quantity before and after impact,
\begin{align}
\Delta g (r,h) = 2\pi r \left( g_\text{after}(r,h) - g_\text{before}(r,h) \right),
\label{17}
\end{align}
allows us to monitor the areas of the film where creation and annihilation of $sp^3$ bonds
take place. The results of this analysis for all the deposition energies studied are shown
in \fig{16}. The figure shows heat maps for $\Delta g(r,h)$ averaged over the last
4000 deposition events at each deposition energy. From this figure, we can infer how at
low energies the rebonding processes take
place in the immediate vicinity of the impact site. At these low energies, rebonding
statistics in the bulk of the film (away from the impact site) are noisy 
due to regular thermal fluctuations. However, as the deposition energy enters the ta-C regime, 
at and beyond 20~eV, we see a clear pattern where $sp^2$ is formed around the impact site
but $sp^3$ bonds are formed \textit{laterally and away} from the impact site. In particular,
\fig{16} shows this as the transition from the noisy heat maps at low energies into solid
net local $sp^3$ density increases at higher deposition energies. It is also interesting to
see that at high deposition energies there is a clear local \textit{annihilation} of
$sp^3$-bonded atoms within an impact cylinder approximately 4~{\AA} wide and 10~{\AA} deep.
This observation
is incompatible with the subplantation model. Instead, at high energies $sp^3$ carbon
is created over a wide film region surrounding this impact cylinder. At low energies,
the incident carbon atoms simply become attached to the surface, where $sp$ sites
offer favorable conditions for adsorption~\cite{caro_2018}. Hence, we propose to
call the low energy process ``direct attachment'', in contrast to the high-energy
mechanism. Both growth processes are schematically depicted in \fig{growth_mechanism_sketches}.

We would like to highlight again that, in view of the large number of bond creation and
annihilation events per impact (\fig{15}), it is remarkable that the GAP succeeds at
correctly describing the extremely delicate balance between $sp^2$ and $sp^3$ creation
(\fig{16}) that leads to the growth of diamond-like a-C at high deposition energies.

\subsection{Elastic properties}

\begin{figure}[t]
\includegraphics{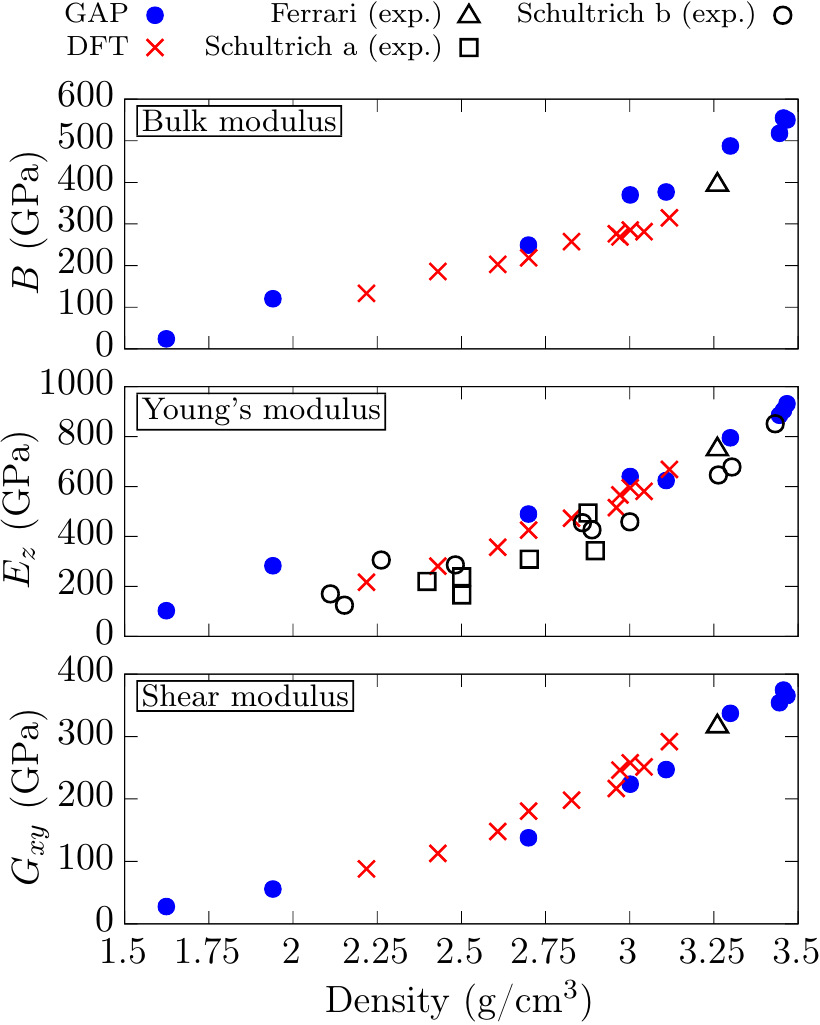}
\caption{Elastic properties of ta-C, as dependent on the mass density of the samples, obtained from
our GAP simulations (blue) and compared to experimental benchmarks where available (open symbols).
Results from DFT, obtained using an indirect (i.e., not deposition) generation
method~\cite{laurila_2017,caro_2014}, are also given (red).}
\label{09}
\end{figure}

Understanding the elastic properties of a-C is particularly important since the main
industrial applications of ta-C coatings relate to friction and wear.
Academically, the elastic properties of diamond-like materials are interesting too,
since diamond itself is (to date) the hardest known material. To understand how
the elastic properties of a-C evolve with mass density, we applied the methodology discussed in
Sec.~\ref{18} to compute elastic moduli for our structures. The results
for bulk modulus, Young's modulus and shear modulus, as a function of density, are shown
in \fig{09} and compared there to experiment and previous DFT results. Overall, very good
agreement with the limited experimental data is observed. All elastic moduli of
a-C increase rapidly as a function of density. Surprisingly, the highest-density samples
show bulk moduli $B$ in excess of the bulk modulus of pure diamond (442~GPa), suggesting that
superhard
ta-C should be less compressible than diamond. 
On the other hand, the Young's modulus $E$ and shear modulus of all the
computational samples stays well below the values of diamond (1053 GPa and 578~GPa,
respectively).

The important result that ta-C is predicted to be less compressible than pure diamond deserves further
attention. While there are many experimental data points
for Young's modulus available from the literature, 
we could only find one
experimental measurement for the bulk modulus, from Ferrari {\etal}~\cite{ferrari_2000}.
Yet, a detailed analysis of that one experiment allows us
to better understand the elastic properties of ta-C and put our results into context.
We give this analysis, together with a discussion on the symmetry of the stiffness
tensor of deposited a-C, in Appendix~\ref{appendix-bulk-modulus}.

\begin{table*}[t]
\caption{Elastic properties of as-deposited a-C films, computed using GAP as described
in Sec.~\ref{18}. The diamond values are provided for comparison.}
\begin{ruledtabular}
\begin{tabular}{l r r r r r r r r}
& 1~eV & 2~eV & 5~eV & 10~eV & 20~eV & 60~eV & 100~eV & Diamond (exp.~\cite{mcskimin_1972})
\\ \hline
In-plane stress (GPa) & $-1.2$ & $0.0$ & $-5.2$ & $-0.3$ & $-9.1$ & $-12.4$ & $-6.5$ & n/a
\\
Out-of-plane stress (GPa) & $-0.2$ & 1.2 & 1.0 & $-0.1$ & $-1.3$ & $-0.3$ & $2.5$ & n/a
\\[0.5em]
$C_{11}$ (GPa) & 52 & 203 & 415 & 922 & 1068 & 1050 & 989 & 1079
\\
$C_{12}$ (GPa) & $-3$ & 91 & 139 & 248 & 319 & 319 & 280 & 124
\\
$C_{13}$ (GPa) & 16 & 56 & 156 & 280 & 297 & 290 & 279 & 124
\\
$C_{33}$ (GPa) & 113 & 305 & 578 & 929 & 1032 & 1054 & 1008 & 1079
\\[0.5em]
$B$ (GPa) & 24 & 123 & 250 & 488 & 555 & 550 & 518 & 442
\\
$E_z$ (GPa) & 103 & 283 & 490 & 795 & 905 & 931 & 885 & 1053
\\
$G_{xy}$ (GPa) & 27 & 72 & 127 & 324 & 363 & 361 & 356 & 578
\end{tabular}
\end{ruledtabular}
\label{19}
\end{table*}

Built-in stresses and detailed elastic properties are given in Table~\ref{19}. As has been 
discussed in the literature, we observe large built-in in-plane compressive stresses in the
high-density films, whereas the out-of-plane stresses are smaller and can be compressive
or tensile. Together with the large differences between $C_{11}$ and $C_{33}$, on the one hand,
and $C_{12}$ and $C_{13}$, on the other, this is a clear indication of film anisotropy. The
role of compressive stresses merits further
discussion since it has been debated in the literature whether these large stresses are necessary
for ta-C growth. In this context, built-in stress can have either of two natures: ``primary'' or
``secondary''. We define primary stress as a necessary condition for high $sp^3$ ta-C growth to
occur, whereas we define secondary stress as the \textit{consequence} of how growth occurs. As can
be seen from our data, all of the superhard ta-C samples (20~eV and above) show very large
built-in stresses of around $-10$~GPa. However, there is a ta-C sample with small built-in
compressive stress, the 10~eV one, which shows a high $sp^3$ fraction of $\sim 82$~\%. Therefore,
on this basis, but keeping in mind the lack of a dataset comprehensive enough to draw
stronger conclusions, we speculate that high compressive stresses in ta-C are indeed
secondary in nature. That is, we speculate that they are a consequence of how ta-C growth
takes place but not a necessary condition for high $sp^3$ fractions to occur.

\section{Conclusions}

ML-driven deposition simulations, mimicking the impact of individual atoms on a surface
at close-to-DFT accuracy, have been shown to be a powerful method for describing and
understanding the properties of amorphous carbon materials.
While our initial contribution dealt with dense ta-C films \cite{caro_2018}, here we have
outlined a more general methodology that also describes low-density forms.
The growth mechanism is strongly dependent on the impacting atom's energy (as is the resulting
structure); at high energies, our simulations suggest peening to be the dominant mechanism
\cite{caro_2018}, whereas at low energies, the simulated films grow by direct formation of $sp$ and $sp^{2}$
motifs around the impact site (\fig{growth_mechanism_sketches}).
We carried out a comprehensive study of structural and mechanical properties, which is in good
agreement with existing experiments and could help with the planning and interpretation of new
ones. The structural models presented here can enable further studies of amorphous carbon
materials for diverse technological applications, 
including friction management \cite{Erdemir2006, Kunze2014, Ma2014, erdemir2018}, 
batteries \cite{deringer_2018a, Yang2011}, 
or biomedical sensing \cite{Zeng2014, laurila_2017, Mynttinen2019, Triroj2020}. 
The predicted formation of $sp^{2}$-rich structures at low impact energies, and the suggestion
of a finely tuned balance between competing coordination environments by varying the energy of
the impacting ions, may in the future be tested by experiments.
The computational approach, making use of fast and accurate ML potentials, 
appears to be promising for predictive studies of other amorphous functional materials.

\begin{acknowledgments}
M.A.C. acknowledges personal funding from the Academy of Finland under projects \#310574
and \#330488.
V.L.D. acknowledges a Leverhulme Early Career Fellowship.
Parts of this work were carried out during V.L.D.'s previous affiliation with the University 
of Cambridge (until August 2019) with additional support from the Isaac Newton Trust.
Parts of this work have been supported by the Project HPC-Europa3 (INFRAIA-2016-1-730897),
with the support of the EC Research Innovation Action under the H2020 Programme.
The authors acknowledge CSC -- IT Center for Science, Finland, 
for computational resources. The authors thank N.~A. Marks for bringing the issue of 5-fold
coordinated atoms to their attention, as well as for stimulating discussions on interatomic
potential simulation of carbon.
\end{acknowledgments}

\appendix

\section{Performance of different empirical force fields for a-C geometries}\label{23}

\begin{figure*}[t]
\includegraphics[width=\textwidth]{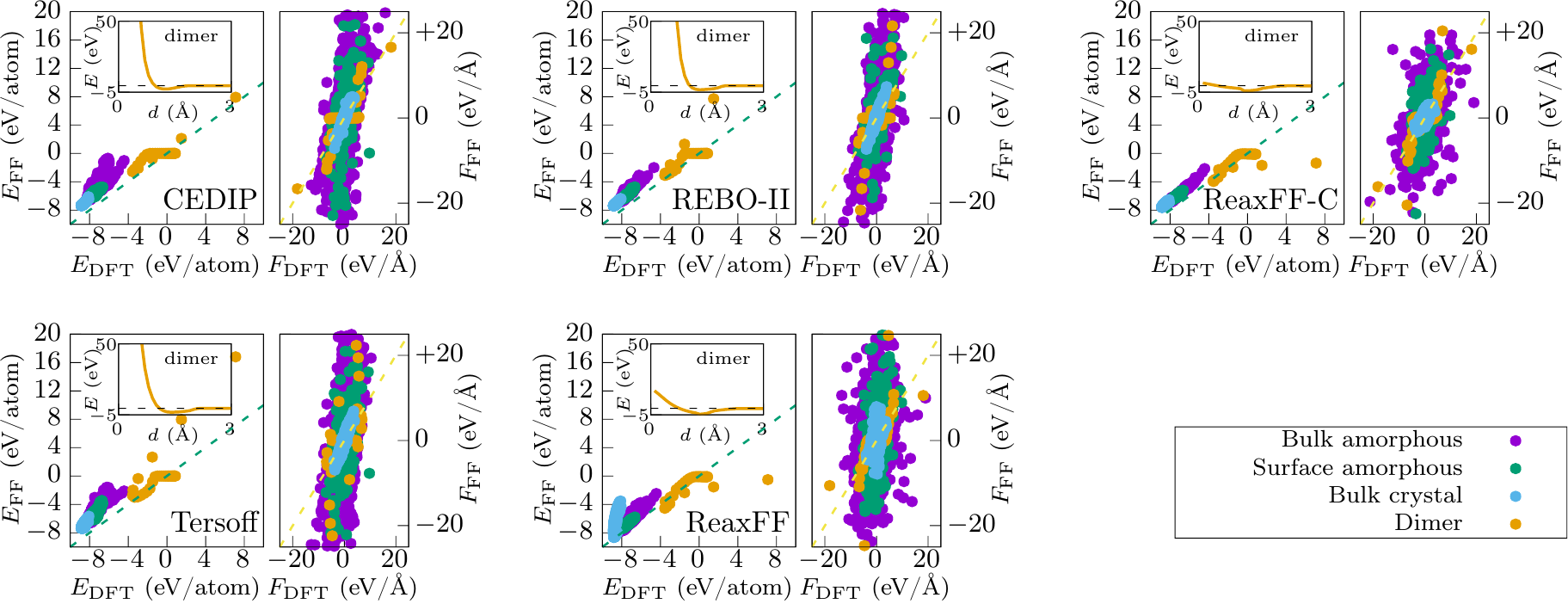}
\caption{Performance of different existing, empirically fitted,
force fields for carbon on the database of atomic structures
used to train the GAP used in this work~\cite{deringer_2017}. 
Note that the main part of this database, labeled as ``bulk amorphous'' for simplicity,
also includes (very) high temperature liquid phases and thus contains structures with rather
high overall energies.
The predictions of each force field
are compared to those of DFT. In the inset, we show the prediction for the C dimer binding curve.
Both versions of ReaxFF lack exchange repulsion, to which their inability to
carry out accurate deposition simulations can be traced back,
as discussed in Sec.~\ref{27} (see also \fig{06}).}
\label{24}
\end{figure*}

To complement the discussion of Sec.~\ref{27} and \fig{06} comparing different force fields
for deposition simulations, in \fig{24} we show the predictions of each of those for the
reference database of the a-C GAP that we use~\cite{deringer_2017}. The most important
point to notice is that the main criterion determining the suitability of the different force
fields for deposition simulations seems to be how accurately the exchange repulsion interaction
is represented. Indeed, \fig{24} shows that even though ReaxFF-C~\cite{srinivasan_2015}
outperforms the other force fields for crystal, surface and bulk amorphous structures, the
energetics of the dimer interaction at short interatomic distances is misrepresented. On the
other hand, CEDIP, which performs best among these classical potentials for deposition
(but still worse than GAP), does a very good job at reproducing the DFT prediction
for the dimer interaction. An attempt to correct the unphysical behavior at small interatomic
distances for ReaxFF-C has been made by Yoon \etal~\cite{yoon_2016}. We also tested that force
field, but unfortunately it did not improve upon the results of ReaxFF-C, because the
short-range repulsion (which we characterized using an isolated C$_2$ molecule as a proxy) is limited to
extremely short interatomic distances. For instance, the ReaxFF-C dimer curve shown here and that
computed with the revised version from Ref.~\onlinecite{yoon_2016} are almost identical up to 0.3~{\AA}
interatomic separation, beyond which the revised force field grows very steeply. As a consequence of
this repulsive behavior (or partial lack thereof), a high-energy incident
atom will be able to get very close to other atoms in the growing film without losing
much of its energy, and will therefore penetrate extremely deep into the film.

\section{Experimental bulk modulus of ta-C}\label{appendix-bulk-modulus}

\begin{figure}[t]
\includegraphics{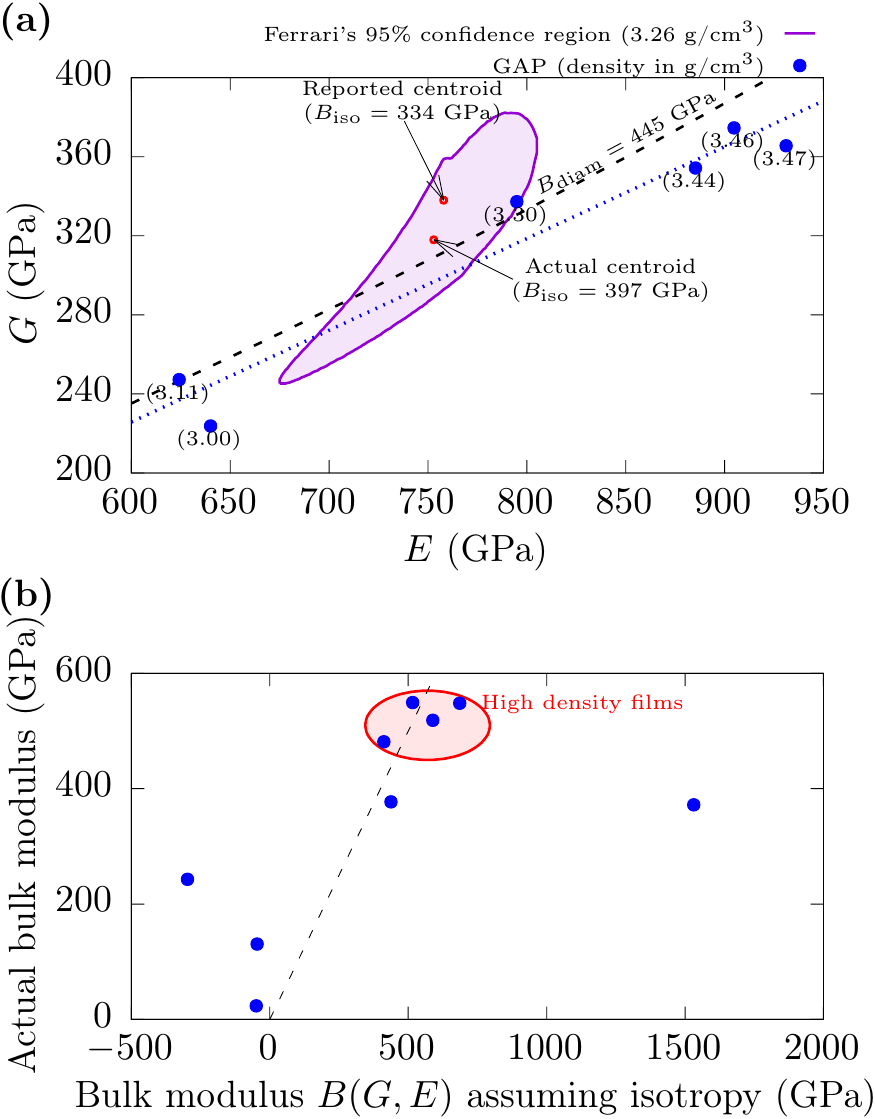}
\caption{(a) Elastic properties of ta-C films as visualized in the $G-E$ plane, including an
experimentally derived 95\% confidence region from Ferrari {\etal}~\cite{ferrari_2000}
and computed values
obtained by GAP for our dense deposited films. The bulk modulus line for diamond is
obtained by rearranging \eq{eq:B} into $G = 3BE / (9B-E)$ and plotting $G(E)$ at a given
constant $B$; note that
it assumes isotropy, which is a simplification (see text for discussion), and highly
inaccurate for the lower density films, as shown in panel (b).}
\label{10}
\end{figure}

The experimental method used by Ferrari {\etal}~\cite{ferrari_2000}
is a type of surface acoustic wave (SAW)
technique known as ``surface Brillouin scattering'' (SBS). Within SBS, the Young's modulus $E$ and
the shear modulus $G$ are obtained \textit{simultaneously}, with a certain degree of confidence. In
particular, Ferrari {\etal} report a 95\% confidence region in the $E$ vs. $G$ plot, as shown in
\fig{10}. From the most
likely pair of values within this region, usually taken as the region's centroid, one can
estimate the corresponding bulk modulus of an equivalent \textit{isotropic} material, since
isotropic materials only have two independent elastic moduli:
\begin{align}
\label{eq:B}
B = \frac{EG}{9G-3E}.
\end{align}

Ferrari's result, together with
our values for 6.5, 8, 10, 20, 60 and 100~eV depositions, are shown in \fig{10}~(a). The bulk modulus reported
in Ref.~\onlinecite{ferrari_2000}, $B = 334$~GPa, was obtained by discarding
(somewhat arbitrarily) 
the portion of the 95\% confidence
region that corresponds to bulk moduli larger than that of pure diamond ($B = 445$~GPa). We
also mark on the figure the position of the centroid of the {\em full} 95\% confidence region (including the $B > 445$~GPa region).
This centroid's coordinates
$(E_\text{c}, G_\text{c})$ were computed as
\begin{align}
E_\text{c} = \frac{\int_{S_{95}} E \, \text{d}G \, \text{d}E}{\int_{S_{95}} \text{d}G \, \text{d}E} 
\quad \text{and} \quad
G_\text{c} = \frac{\int_{S_{95}} G \, \text{d}G \, \text{d}E}{\int_{S_{95}} \text{d}G \, \text{d}E},
\end{align}
where $S_{95}$ denotes the region of 95\% confidence over which the integrals extend.
In this case, one obtains the centroid shown which
gives the bulk modulus that best fits Ferrari's data, $B = 397$~GPa, \textit{assuming elastic
isotropy}. As shown in \fig{10}~(b), the assumption of isotropy for a-C film is particularly
bad for low density films, which are highly oriented along the growth axis.

Grown a-C
films are not isotropic since the growth direction is clearly singled out and therefore breaks
the material's symmetry. This is further supported by the fact that large in-plane stresses exist in
as-grown a-C films. Therefore, instead of an
isotropic stiffness tensor $\mathbb{C}_\text{iso}$:
\begin{align}
& \mathbb{C}_\text{iso} = \left(
\begin{array}{cccccc}
C_{11} & C_{12} & C_{12} & 0 & 0 & 0 \\
C_{12} & C_{11} & C_{12} & 0 & 0 & 0 \\
C_{12} & C_{12} & C_{11} & 0 & 0 & 0 \\
0 & 0 & 0 & \frac{C_{11}-C_{12}}{2} & 0 & 0 \\
0 & 0 & 0 & 0 & \frac{C_{11}-C_{12}}{2} & 0 \\
0 & 0 & 0 & 0 & 0 & \frac{C_{11}-C_{12}}{2}
\end{array}
\right),
\end{align}
our calculated values are obtained allowing for a lower symmetry
stiffness tensor $\mathbb{C}_\text{hex}$, corresponding to hexagonal symmetry,
\begin{align}
& \mathbb{C}_\text{hex} = \left(
\begin{array}{cccccc}
C_{11} & C_{12} & C_{13} & 0 & 0 & 0 \\
C_{12} & C_{11} & C_{13} & 0 & 0 & 0 \\
C_{13} & C_{13} & C_{33} & 0 & 0 & 0 \\
0 & 0 & 0 & C_{44} & 0 & 0 \\
0 & 0 & 0 & 0 & C_{44} & 0 \\
0 & 0 & 0 & 0 & 0 & \frac{C_{11}-C_{12}}{2}
\end{array}
\right),
\end{align}
which preserves transverse isotropy. Note that while the symmetry of the stiffness tensor
of these films is hexagonal in the limit of infinite system size, the actual simulation
cells are themselves orthorhombic.

The condition of elastic isotropy is given by the relation between the shear elastic constants
and the axial elastic constants, $C_{44} = C_{55} = C_{66} = \frac{1}{2} (C_{11}-C_{12})$. 
Therefore,
for our film the in-plane symmetry is preserved by the use of the hexagonal stiffness tensor, which
presents in-plane isotropy, $C_{66} = \frac{1}{2} (C_{11}-C_{12})$~\footnote{Note that, numerically,
$C_{11}$ is obtained by averaging $C_{11}$ and $C_{22}$, which differ due to the finite size
of our system.}. Generalized expressions for bulk, Young's and shear moduli of these films, which
explicitly incorporate the correct underlying symmetry of the films, are:
\begin{align}
& B \approx \frac{2C_{11} + C_{33} + 2 C_{12} + 4 C_{13}}{9} \qquad \text{(hydrostatic strain)},
\nonumber \\
& B = \frac{(C_{11}+ C_{12}) C_{33} -2 {C_{13}}^2}{C_{11} + C_{12} - 4 C_{13} + 2 C_{33}} \qquad \text{(hydrostatic stress)},
\nonumber \\
& E_z = C_{33} - \frac{2{C_{13}}^2}{C_{11} + C_{12}} \qquad \text{(along growth axis $z$)},
\nonumber \\
& G_{xy} = \frac{C_{11} - C_{12}}{2} \qquad \text{(in growth plane $xy$)},
\end{align}
where the bulk modulus can be computed assuming deformation under applied hydrostatic strain
(approximately correct for quasi-isotropic materials) or deformation under applied hydrostatic
stress (always correct).

\end{document}